\pdfoutput=1
\documentclass[11pt]{article}
\usepackage{amsmath,amssymb,mathtools}
\usepackage{natbib}
\usepackage[font=small]{caption}
\usepackage{enumitem}
\usepackage{bm}
\usepackage{ying}
\usepackage{multirow}
\usepackage{authblk}
\usepackage{geometry}
\usepackage{parskip}
\allowdisplaybreaks
\usepackage{float}
\usepackage{subcaption}
\usepackage{tabularx}
\PassOptionsToPackage{dvipsnames}{xcolor}
\usepackage{graphicx}
\usepackage{xcolor}
\usepackage[colorlinks,
linkcolor=red,
anchorcolor=blue,
citecolor=blue
]{hyperref}
\hypersetup{pdftitle={A uniform improvement to conformal risk control via a generalization of e-closure},
	pdfauthor={Tian Bai and Aaditya Ramdas}}

\providecommand{\keywords}[1]
{
	\fontsize{9}{12}\selectfont
	\textbf{\textit{Keywords:}} #1
}

\newtheorem{proposition}[theorem]{Proposition}
\numberwithin{equation}{section}
\makeatletter
\renewcommand\paragraph{\@startsection{paragraph}{4}{\z@}%
	{3.25ex \@plus1ex \@minus.2ex}{-1em}{\normalfont\fontsize{10}{12}\selectfont\bfseries}}
\makeatother

\newcommand{\Prob}{\mathbb P}

\newcommand{\Sc}{\mathcal S}
\newcommand{\Ac}{\mathcal A}
\newcommand{\Xc}{\mathcal X}
\newcommand{\Zc}{\mathcal Z}
\newcommand{\Pc}{\mathcal P}
\newcommand{\Mc}{\mathcal M}
\newcommand{\Cc}{\mathcal C}

\newcommand{\FDP}{\operatorname{FDP}}

\newcommand{\CRC}{A_{\mathrm{CRC}}}
\newcommand{\ENV}{A_{\mathrm{env}}}
\newcommand{\rec}{\mathcal R}

\DeclareMathOperator{\eBH}{eBH}

\title{A uniform improvement to conformal risk control\\via a generalization of e-closure}
\author{Tian Bai}
\author{Aaditya Ramdas}
\affil{Stanford University\\ \texttt{\{tianbai,aramdas\}@stanford.edu}}
\date{}

\begin{document}

	\maketitle

	\begin{abstract}
		\fontsize{9}{11}\selectfont
		We extend the recently introduced e-closure principle for multiple testing to general state and action spaces and to arbitrary, including continuously indexed, families of risks. A collection of e-values indexed by states and risks certifies a menu of actions, and every data-dependent selection from this menu controls each expected loss at its prescribed level. Conversely, every valid procedure is certified by its own normalized losses, which often leave part of the risk budget unused. We introduce a method called \emph{recertification}, which spends this unused budget. Using known distributional structure, it enlarges the certifying e-values while preserving validity and selects a less conservative action from the enlarged menu.
		One particularly compelling application is to the popular framework of conformal risk control (CRC) with a known loss family. Here, recertification redistributes unused risk budget among the held-out points of each hypothetical augmented sample. The resulting controller is valid under every finite exchangeable law and less (usually strictly) conservative than CRC. For convex losses on an interval of actions, repeated recertification converges geometrically to an admissible controller.
		As a second example involving false discovery rate control with e-values, recertifying randomized e-BH yields a closed-form procedure that rejects a superset of randomized e-BH for every realization of the randomization and maximizes the expected number of rejections under the same constraints. Recertifying with the larger e-collection of closed e-BH gives a second closed-form procedure that rejects a superset of the first.
		Experiments on CIFAR-100 hierarchical classification and simulations demonstrate improved budget utilization and gains in informativeness and power over the corresponding baselines.
	\end{abstract}
	\fontsize{10}{12}\selectfont

	\section{Introduction}\label{sec:intro}
	Many statistical procedures turn data into an informed action, while guaranteeing that some expected loss stays below a prescribed level. Statistical decision theory formalizes this task through a decision rule mapping data to actions, a loss depending on the action and an unknown state of nature, and the resulting statewise expected loss, or risk \citep{wald1950,berger1985}. The bound imposes a risk budget that must hold in every state. Among rules satisfying this constraint, one seeks the most useful decisions, as in the constrained problems of \citet{hodgeslehmann1952}.

	Multiple testing is a decision problem of exactly this kind. With $K$ hypotheses, the action is a rejection set $R\subseteq[K]$, so the action space is the collection $2^{[K]}$ of all subsets of $[K]$, and the unknown state is the set $T\subseteq[K]$ of true null hypotheses. The false discovery proportion $|T\cap R|/(|R|\vee1)$ is the loss, and false discovery rate (FDR) control bounds its expectation by $\alpha$. The procedure of \citet{benjaminihochberg1995} achieves this by comparing the ordered p-values with linearly increasing thresholds; its e-value analogue, the e-BH procedure of \citet{wangramdas2022}, applies the reciprocal thresholds to e-values, which are nonnegative statistics with expectation at most one under their null hypotheses \citep{vovkwang2021,ramdaswang2025}.
	The recent work of \citet{xu2026closure} introduces the e-closure principle, a general principle that uniformly improves many testing procedures based on e-values, including e-BH. Specifically, given an e-value $E_T$ for each candidate true-null set $T$, valid when $T$ is the set of true nulls, the principle certifies every rejection set $R$ that satisfies
	\begin{equation}\label{eq:intro-fdr}
		\frac{|T\cap R|}{|R|\vee1}\leq\alpha E_T\qquad\text{for every }T\subseteq[K].
	\end{equation}
	Under the actual null set, the left side is the false discovery proportion and the right side has expectation at most $\alpha$. The false discovery rate is therefore controlled even when the analyst chooses among the certified rejection sets after seeing the data. Conversely, \citet{xu2026closure} show that a procedure controls the false discovery rate simultaneously over the rejection sets it may report if and only if all of them are certified in this way by some family of e-values. The principle applies equally to other error rates that are expected losses, and it contains the classical closed testing principle for the familywise error rate \citep{marcus1976} as the case of e-values taking only the values $0$ and $1/\alpha$.

	Our first contribution is to note that nothing in this argument depends on the actions being rejection sets, so e-closure extends well beyond multiple testing. In a general decision problem, a state $s$ specifies a class of distributions for the data $X$, an action $a$ is chosen from an arbitrary set, and each of several losses $L_t(s,a;X)$, indexed by $t$, carries its own budget $\alpha_t(s)$. A family of e-values, one for each state and risk, certifies the actions whose losses fit within all the budgets it allows. Every data-dependent choice among the certified actions then controls every risk, and every procedure with this guarantee is certified by some such family \citep{grunwald2023eposterior,grunwald2024beyond,andrewschen2025}. In Section~\ref{sec:closure} we develop the e-closure principle at this level of generality, with menus of arbitrary actions, possibly continuously indexed states and risks, and losses that may depend on the data.

	The correspondence can be read in both directions, and the direction from procedures to e-values suggests how to improve a procedure. Improving a decision rule directly is difficult: a more liberal rule may have higher power or utility, but it is rarely clear how far a given rule can be pushed before its guarantee fails. By contrast, improving an e-value is easier. If an e-value has expectation strictly below one, and the shortfall can be computed, for instance conditionally on a statistic whose conditional law is known under the null, then adding the shortfall back yields a larger e-value.
	Through the e-closure correspondence, larger e-values translate back into decisions.
	Starting from a valid procedure, we apply this construction to its canonical e-values: its losses divided by their risk budgets. Inverting the enlarged certificate yields a menu containing the original actions and potentially more liberal ones, all with the same guarantee. We call this step \emph{recertification}. It generalizes the anchoring idea of \citet{Ramdas2026Anchored}, which keeps an existing false discovery rate procedure and spends the slack in its validity proof, and it can be repeated.

	We apply recertification to two problems. The first is conformal risk control (CRC) \citep{angelopoulos2024crc}, which extends conformal prediction from coverage to general monotone losses. Given $n$ exchangeable calibration points and a known family of losses indexed by a conservativeness parameter $\lambda$, CRC selects the least conservative $\lambda$ for which the calibration losses, together with a bound on the loss of the unseen test point, fit within the budget $(n+1)\alpha$. The selected controller typically leaves part of this budget unused.
	Recertification computes this unused budget on every such sample, redistributes it among the $n+1$ held-out positions, and inverts the resulting e-values. The new controller is valid under every finite exchangeable law, and for every calibration sample it is no more conservative than CRC. In this sense, for a fixed loss family and observation domain, it improves CRC uniformly. This recertification step can be repeated, with diminishing returns. For convex losses, the iterates converge geometrically to a controller that admits no further pointwise improvement, and on finite outcome spaces a coordinatewise sweep reaches such a controller without convexity (Section~\ref{sec:crc}).

	\begin{figure}[!htbp]
		\centering
		\includegraphics[width=1\textwidth]{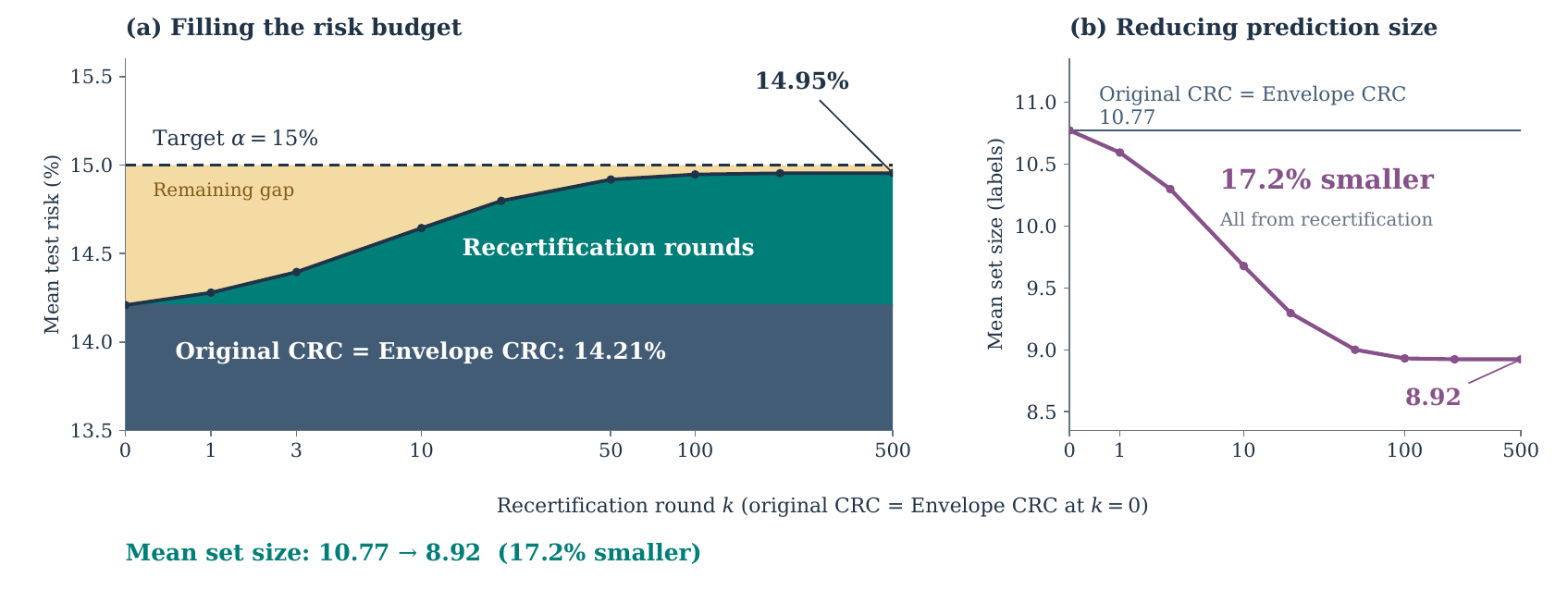}
		\caption{Risk-budget use and mean prediction-set size on CIFAR-100 with confidence-adaptive prediction rules ($n=50$, $\alpha=0.15$). Round $0$ is CRC, and later rounds apply recertification. Results are averaged over 300 paired splits, each with 16 test images.}
		\label{fig:intro_preview}
	\end{figure}

	Figure~\ref{fig:intro_preview} illustrates these gains in hierarchical classification on CIFAR-100 (Section~\ref{sec:sim}), using rules that give more specific predictions when the classifier is more confident. At a target of $15\%$, CRC attains a mean test risk of $14.21\%$. Repeated recertification raises it to $14.95\%$, close to the target, and reduces the mean prediction-set size from $10.77$ to $8.92$ labels, a reduction of $17.2\%$.

	The second application returns to multiple testing. The e-BH procedure is valid under arbitrary dependence but can be conservative, and \citet{xu2023randomization} showed that external randomization recovers part of the lost power. Their U-eBH procedure boosts the e-values by dividing them by a single uniform random variable and then applies e-BH.
	In this setting, we treat the conditional law of the rejection set given the e-values as a randomized action. In these terms, U-eBH is certified by the family of e-values behind e-BH, through a version of the self-consistency condition of \citet{blanchard2008two} that holds on average over the randomization, and it typically leaves part of the budget allowed by this family unused. Recertification spends this budget and yields a closed-form randomized rule, recertified U-eBH (RU-eBH), which rejects at least as many hypotheses as U-eBH for every realization of the randomization and has the largest expected number of rejections among all randomized procedures certified by the same family.

	The same principle applies to closed e-BH, the improvement of e-BH by e-closure \citep{xu2026closure}, whose family of e-values is larger because it adapts to the number of true nulls. Recertifying with this family gives a second closed-form rule, recertified closed e-BH (RC-eBH), whose rejection set contains that of RU-eBH and which has the largest expected number of rejections among all randomized procedures certified by this family, closed e-BH included (Section~\ref{sec:rebh}). In simulations, both rules are more powerful than e-BH and U-eBH in every setting, and RC-eBH is the most powerful of the eight procedures we compare in nearly every setting.

	\subsection{Related work}\label{subsec:related}
	\paragraph{E-values and decisions.}
	E-values have become a standard tool for testing and multiple testing; they remain valid when the significance level is chosen after seeing the data \citep[Chapter~4]{ramdaswang2025}. This property was identified by \citet{wangramdas2022}, \citet{katsevich2020simultaneous} and \citet{xu2024postselection} in multiple testing, and in particular when testing a single hypothesis.
	\citet{grunwald2024beyond} developed this into a generalization of the Neyman--Pearson framework to data-dependent levels in an expected-loss setting. \citet{koning2023posthoc} then showed that e-values are necessary for post hoc testing and introduced the post hoc p-value, \citet{chugg2026admissibility} studied admissibility in this framework, and \citet{koning2026axioms} derived its validity criterion from axioms.
	For a single risk, our certified menu is the set of actions whose e-posterior risk bound \citep{grunwald2023eposterior}, with each state's loss divided by its budget, is at most one.
	\citet{andrewschen2025} study decisions reported together with certificates of their loss, and \citet{koningvanmeer2026} give evidence-weighted loss bounds for decisions based on conformal e-prediction sets. Decision-theoretic views of conformal prediction also study how prediction sets should inform downstream actions \citep{lekeufack2024conformal,kiyani2025decision}. These works allow general states and actions; our focus is the constructive use of the correspondence, which keeps the error budget fixed and recertifies an existing procedure.

	\paragraph{Closure principles.}
	Closed testing \citep{marcus1976} controls the familywise error rate, and every coherent procedure for that error rate is a closed testing procedure \citep{sonnemann2008}. Closed testing also yields simultaneous bounds on the false discovery proportion of every rejection set \citep{goemansolari2011}, and only closed testing procedures are admissible for that purpose \citep{goeman2021only}. E-closure \citep{xu2026closure} extends the closure principle to error rates defined by expected losses. Its follow-ups include the uniform improvement of the Benjamini--Hochberg procedure \citep{Goeman2026Closed}, anchoring \citep{Ramdas2026Anchored}, online versions \citep{xu2026online}, complete-class and admissibility results for e-value procedures \citep{sunwang2026}, and a continuous closure for allocating resources among strategic agents through statistical contracts \citep{chugg2026strategic}.
	\citet{koning2026emeasure} closes evidence over classes of hypotheses through e-measures, which yield evidence bounds holding uniformly over data-informed decisions and connect to e-closure in multiple testing. We note that representation and admissibility are separate questions. Every valid procedure is certified by some family of e-values, while admissibility requires further conditions, and our admissibility results for CRC rely on the structure of that problem.

	\paragraph{Randomized multiple testing.}
	Randomization is classical in testing, where it makes tests based on discrete statistics exact \citep{lehmannromano2022}. For false discovery rate control with e-values under arbitrary dependence, \citet{xu2023randomization} showed that stochastic rounding of the e-values and U-eBH increase the power of e-BH, and \citet{xu2026closure} combine stochastic rounding with e-closure. These methods randomize the e-values. Our procedures instead choose the distribution of the rejection set directly, subject either to per-hypothesis budgets that form a randomized version of the self-consistency condition of \citet{blanchard2008two} or to the constraints that certify closed e-BH. Other uniform improvements of e-BH include its minimally adaptive version \citep{ignatiadis2024evalues}, closed e-BH \citep{xu2026closure}, and conditional calibration \citep{fithianlei2022,leeren2024}, which recalibrates the e-values given a sufficient statistic.
	All of these improvements fit our framework. The first two supply larger certifying e-collections, and recertifying with them using randomization improves upon both (Section~\ref{subsec:rebh-mean} and Appendix~\ref{app:rebh-ma}); conditional calibration is itself a form of recertification (Remark~\ref{rem:cc}).

	\paragraph{Conformal risk control.}
	CRC already handles several, even infinitely many, risks that share a monotone action parameter \citep[Section~4.3]{angelopoulos2024crc}. Risk-Controlling Prediction Sets (RCPS) and Learn then Test (LTT) instead provide high-probability statements about population risks \citep{bates2021rcps,angelopoulos2025ltt}. Extensions of CRC treat non-exchangeable data \citep{farinhas2024nonexchangeable}, cross-validation \citep{cohen2024cross}, non-monotonic losses \citep{angelopoulos2026nonmonotonic}, and end-to-end training \citep{yeh2025training}. The leave-one-out characterization of validity that we use (Lemma~\ref{lem:orbit}) is the risk-control counterpart of the universality of conformal prediction \citep[Proposition~2.9]{vovk2022algorithmic}; see also \citet{tibshirani2026lens}, and \citet{angelopoulos2026nonmonotonic} for its sufficiency direction. Conformal prediction with e-values is developed by \citet{vovk2025eprediction}, \citet{koning2026invariance}, \citet{gauthier2025evalues}, and \citet{koningvanmeer2026}. Apart from the sharper bound on the unseen test loss in envelope CRC \citep{yeh2025training}, which serves as our starting point, we are not aware of work on improving CRC controllers pointwise or on their admissibility. We use CRC to show concretely how large such improvements can be and what they cost in computation.

	\subsection{Organization}
	Section~\ref{sec:closure} formulates e-closure for general decision problems, including randomized actions, and introduces recertification. Section~\ref{sec:crc} develops the improvement of CRC, including its admissibility theory, worked examples and a real-data study. Section~\ref{sec:rebh} returns to multiple testing, recertifies randomized e-BH with the e-collections of e-BH and closed e-BH, and reports simulations on Gaussian mean testing. Section~\ref{sec:utility} comments on the choice among certified actions and concludes.

	\section{Generalized e-closure and recertification}\label{sec:closure}
	This section sets up the decision framework used throughout the paper. We first describe states, actions and losses, then establish the correspondence between e-values and valid decisions, and finally introduce recertification, which turns this correspondence into a way of improving existing procedures.

	\subsection{States, actions, and a family of risks}
	Suppose that a statistician observes data $X$ taking values in a space $\Xc$. The distribution of $X$ is governed by an unknown state $s\in\Sc$, and each state specifies a nonempty class $\Pc_s$ of distributions of $X$. After observing $X$, the statistician takes an action $a\in\Ac$, whose quality is measured by a family of known nonnegative losses
	\[
	L_t(s,a;x),\qquad t\in\mathcal T.
	\]
	The index $t$ ranges over the kinds of error that the statistician wishes to control, such as several error rates or a continuum of loss functions, and each loss may depend on the observation $x$ as well as on the state and the action. The sets $\Sc$, $\Ac$ and $\mathcal T$ are arbitrary and may be uncountable.

	Each loss comes with a budget. Given fixed positive levels $\alpha_t(s)$, a decision rule $A:\Xc\to\Ac$ is \emph{valid} if
	\begin{equation}\label{eq:risk-target}
		\E_P [L_t(s,A(X);X)] \leq\alpha_t(s)
		\quad\text{for every }s\in\Sc,\ P\in\Pc_s,\ t\in\mathcal T.
	\end{equation}
	In words, whatever the state and whichever distribution within it generates the data, the expected loss of the selected action stays within its budget, separately for every loss in the family. The expectation is unconditional: it averages over the data and hence over the action that the rule selects. Throughout, we assume that all decision rules and all losses that we evaluate are measurable, and when we discuss sets of actions we also assume that their statewise loss envelopes, defined below, are measurable. Appendix~\ref{app:measurable} gives sufficient conditions and discusses the uncountable intersections that arise.

	Validity is certified by e-values. Recall that an e-value for a class of distributions is a nonnegative statistic whose expectation is at most one under every distribution in the class. Since \eqref{eq:risk-target} involves every state and every loss, we use one e-value for each such pair. An \emph{e-collection} is a family of nonnegative measurable functions $E_{s,t}$ satisfying
	\begin{equation}\label{eq:earray}
		\E_P E_{s,t}(X)\leq1\qquad(s\in\Sc,\ P\in\Pc_s,\ t\in\mathcal T).
	\end{equation}
	In other words, an e-collection is a set of e-values, one for each combination of state and loss. Because an e-collection will be used to certify actions, we also refer to it as a certificate.

	\subsection{From e-collections to decisions and back}\label{subsec:menus}
	We now make precise the correspondence between e-collections and valid decisions described in the introduction. It is convenient to work with sets of actions rather than with single rules. A \emph{menu} assigns to each observation $x$ a nonempty set $\Mc(x)\subseteq\Ac$ of actions that the statistician is prepared to take. After seeing the data, the statistician may choose any action from $\Mc(X)$, for instance by maximizing a utility or following a scientific preference. A decision rule is the special case of a menu with a single action at each observation. We first show how an e-collection produces a menu, then how a menu produces an e-collection, and finally that the two constructions are compatible.

	\paragraph{From e-collections to menus.}
	An e-collection certifies an action at the observation $x$ when every loss of that action fits within its budget, scaled by the corresponding e-value. The \emph{certified menu} of $E$ is therefore
	\begin{equation}\label{eq:closure}
		\Cc_\alpha(E;x)=\{a\in\Ac:L_t(s,a;x)\leq\alpha_t(s)E_{s,t}(x)\ \forall s,t\}.
	\end{equation}
	The constraint for state $s$ caps the loss an action may incur if $s$ is true, in proportion to $E_{s,t}(x)$. Intuitively, a large e-value indicates that the data are unlikely under $s$, so the menu may include actions that would perform poorly under $s$; a small one restricts the menu to actions that perform well under $s$.

	\paragraph{From menus to e-collections.}
	Conversely, a menu determines the largest loss it can incur in each state. For a menu $\Mc$, define its loss envelopes and their normalized versions by
	\begin{equation}\label{eq:canonical}
		g_{s,t}^{\Mc}(x)=\sup_{a\in\Mc(x)}L_t(s,a;x),
		\qquad E_{s,t}^{\Mc}(x)=\frac{g_{s,t}^{\Mc}(x)}{\alpha_t(s)}.
	\end{equation}
	The menu is valid if $\E_P\,g_{s,t}^{\Mc}(X)\leq\alpha_t(s)$ for every $s\in\Sc$, $P\in\Pc_s$ and $t\in\mathcal T$. This requires \eqref{eq:risk-target} to hold for every selection from the menu simultaneously, and it is equivalent to $E^{\Mc}$ being an e-collection, which we call the \emph{canonical e-collection} of $\Mc$. Applying the first construction to it gives the closure of $\Mc$,
	\begin{equation}\label{eq:lossclosure}
		\overline\Mc(x)=\Cc_\alpha(E^{\Mc};x)=\{a:L_t(s,a;x)\leq g_{s,t}^{\Mc}(x)\ \forall s,t\},
	\end{equation}
	which consists of all actions whose losses never exceed those already incurred by the menu.

	The following theorem, which we call generalized e-closure, states that the two constructions describe the same objects: certified menus are valid, and every valid menu is certified by its canonical e-collection.

	\begin{theorem} \label{thm:closure}
		Let $E$ be an e-collection. Every measurable selection $A(X)\in\Cc_\alpha(E;X)$ satisfies \eqref{eq:risk-target}. More strongly, for every state--loss pair $(s,t)$ and every $P\in\Pc_s$,
		\begin{equation}\label{eq:sim-menu}
			\E_P\left[\sup_{a\in\Cc_\alpha(E;X)}L_t(s,a;X)\right]\leq\alpha_t(s).
		\end{equation}
		Further, a menu $\Mc$ is valid if and only if $\Mc\subseteq\Cc_\alpha(E)$ for some e-collection $E$. The loss closure \eqref{eq:lossclosure} is extensive, monotone under inclusion of menus and idempotent, and it preserves every loss envelope. In particular,
		\begin{equation}\label{eq:closed-canonical}
			\Mc=\Cc_\alpha(E)\quad\Longrightarrow\quad\Mc=\Cc_\alpha(E^{\Mc}).
		\end{equation}
	\end{theorem}
	\begin{proof}
		For every $a\in\Cc_\alpha(E;X)$, definition \eqref{eq:closure} gives $L_t(s,a;X)\leq\alpha_t(s)E_{s,t}(X)$, so the supremum in \eqref{eq:sim-menu} is at most $\alpha_t(s)E_{s,t}(X)$. Taking expectations under $P\in\Pc_s$ and using \eqref{eq:earray} proves \eqref{eq:sim-menu}, and the validity of every measurable selection follows. If $\Mc\subseteq\Cc_\alpha(E)$, then $g^{\Mc}_{s,t}\leq\alpha_t(s)E_{s,t}$, so $\Mc$ is valid. Conversely, if $\Mc$ is valid, then $E^{\Mc}$ is an e-collection, and every action in $\Mc(x)$ satisfies the constraints defining $\Cc_\alpha(E^{\Mc};x)$, so $\Mc\subseteq\overline\Mc$. Every action in $\overline\Mc(x)$ has losses at most $g^{\Mc}_{s,t}(x)$, so $\overline\Mc$ has the same envelopes as $\Mc$; idempotence follows, and monotonicity follows from the monotonicity of the envelopes. Finally, if $\Mc=\Cc_\alpha(E)$, then $\Mc\subseteq\Cc_\alpha(E^{\Mc})$, and $g^{\Mc}_{s,t}\leq\alpha_t(s)E_{s,t}$ gives the reverse inclusion $\Cc_\alpha(E^{\Mc})\subseteq\Cc_\alpha(E)=\Mc$.
	\end{proof}

	By \eqref{eq:sim-menu}, the guarantee holds simultaneously over all actions in a certified menu. The statistician may therefore choose an action after seeing the data, for instance by maximizing a utility or applying a scientific filter, and every such choice from the menu of a fixed e-collection controls every risk. This is closely related to the post hoc validity of e-values in hypothesis testing, where the significance level, and more generally the loss, may be chosen after the data are observed \citep{grunwald2024beyond,koning2023posthoc,chugg2026admissibility}. We also note that the e-values may be arbitrarily dependent, because for each loss the proof uses only the corresponding e-value under the true state.

	\paragraph{Multiple testing as a special case.}
	The e-closure principle for multiple testing is recovered as follows. Take both the state space and the action space to be $2^{[K]}$, let $\Pc_T$ contain the distributions under which $T$ is the set of true null hypotheses, and use the loss $L(T,R)=|T\cap R|/(|R|\vee1)$. The certified menu \eqref{eq:closure} then consists of the rejection sets certified by \eqref{eq:intro-fdr}, which is the e-closure principle for the false discovery rate \citep{xu2026closure}, and the loss $L(T,R)=\ind\{T\cap R\ne\varnothing\}$ gives its familywise error rate version. E-values for intersection hypotheses, which are valid whenever $T$ is contained in the set of true nulls, are in particular valid for the exact-state classes $\Pc_T$ used here. Section~\ref{sec:rebh} reviews this setting in more detail. The general framework requires no structure on the actions, such as unions or an ordering, and no monotonicity of the losses. For instance, the false discovery proportion can decrease when a rejection is added.

	\paragraph{Randomized actions.}
	Actions may also be randomized. A randomized action is a probability measure $Q$ on $\Ac$: once $Q$ is chosen, an action is drawn from it using randomization independent of the data, and the loss of $Q$ is the average $\bar L_t(s,Q;x)=\int L_t(s,a;x)\,Q(da)$. Probability measures on $\Ac$ form an action space like any other, so Theorem~\ref{thm:closure} applies with these averaged losses. Every data-dependent choice of $Q$ from the certified menu $\{Q:\bar L_t(s,Q;x)\leq\alpha_t(s)E_{s,t}(x)\ \forall s,t\}$ is therefore valid, with a guarantee that averages over the subsequent draw; it does not cover choices made after the draw is seen. This menu can contain randomized actions whose possible realizations are not individually certified, so randomization can enlarge what a fixed e-collection certifies. Section~\ref{sec:rebh} uses this to improve randomized multiple testing.

	\subsection{Recertification: spending the unused budget}\label{subsec:recert}
	A valid menu can be enlarged in two ways, and the distinction between them is central to our approach. The first is \emph{closing}, which adds every action whose losses lie beneath the current envelopes of the menu. By Theorem~\ref{thm:closure}, closing preserves every envelope: the closed menu offers more choice, which in multiple testing is the simultaneity gain described by \citet{xu2026closure}, but it uses exactly the budget that the original menu uses. Any budget that the original procedure leaves unused therefore remains unused after closing. The second way, which we call \emph{recertifying}, first replaces the envelopes by larger ones that still respect the budgets and then closes with respect to the new envelopes. The recertified menu can therefore contain actions whose losses exceed those of every action in the original menu, and this is how recertification spends the unused budget.

	To recertify, we need to quantify the budget that the baseline leaves unused. We do so conditionally on a statistic, chosen for each state--loss pair so that the required conditional expectations are known. When the unused budget can be computed without conditioning, the statistic is simply taken to be constant. Fix a valid baseline menu with envelopes $g_{s,t}$, and let $V_{s,t}$ denote the conditioning statistic for the pair $(s,t)$. Suppose that known nonnegative functions $H_{s,t}$ and $B_{s,t}$ satisfy, for every $P\in\Pc_s$,
	\begin{equation}\label{eq:conditional-budget}
		\E_P[g_{s,t}(X)\mid V_{s,t}]\leq H_{s,t}(V_{s,t})\leq B_{s,t}(V_{s,t}),
		\qquad \E_P B_{s,t}(V_{s,t})\leq\alpha_t(s).
	\end{equation}
	The function $H_{s,t}$ bounds the conditional expected loss of the baseline, while $B_{s,t}$ is a conditional budget whose expectation does not exceed the original level. Their difference $B_{s,t}-H_{s,t}$ is therefore an allowance that the baseline leaves unused, conditionally on $V_{s,t}$. The following proposition shows that any nonnegative increment within this allowance can be added to the envelopes.

	\begin{proposition}[Conditional recertification]\label{prop:conditional}
		If measurable functions $W_{s,t}(X)\geq0$ satisfy
		\begin{equation}\label{eq:conditional-increment}
			\E_P[W_{s,t}(X)\mid V_{s,t}]\leq B_{s,t}(V_{s,t})-H_{s,t}(V_{s,t})
			\quad(P\in\Pc_s),
		\end{equation}
		then $(g_{s,t}+W_{s,t})/\alpha_t(s)$ is an e-collection. Its certified menu contains the original menu and every action whose losses are at most $g_{s,t}+W_{s,t}$ for all $s$ and $t$.
	\end{proposition}
	\begin{proof}
		Conditionally on $V_{s,t}$, the expectation of $g_{s,t}+W_{s,t}$ is at most $H_{s,t}+(B_{s,t}-H_{s,t})=B_{s,t}$. Taking expectations once more, the unconditional expectation is at most $\alpha_t(s)$. The two inclusions follow from $W_{s,t}\geq0$ and from the definition \eqref{eq:closure} of the certified menu.
	\end{proof}
	The increments $W_{s,t}$ may depend on the data in an arbitrary way; only their conditional expectations are constrained by \eqref{eq:conditional-increment}. How the allowance is computed depends on the problem. In CRC (Section~\ref{sec:crc}), the state is a hypothetical augmented sample whose law is known by exchangeability, so no further conditioning is needed, and the allowance is shared equally among the held-out points. For randomized e-BH (Section~\ref{sec:rebh}), the input e-values supply the budget directly, and conditional on the e-values, the averaged losses are known because the law of the external randomization is known. Proposition~\ref{prop:conditional} also describes an existing method, the conditional calibration of e-values \citep{fithianlei2022,leeren2024}, which improves e-BH by recalibrating each e-value given a sufficient statistic; there the sufficient statistic provides the conditional expectations (Remark~\ref{rem:cc}). Finally, the recertified e-collection is again an e-collection, so its certified menu is valid by Theorem~\ref{thm:closure} and can serve as the baseline for another round of recertification.

	\section{Improving conformal risk control}\label{sec:crc}
	We now apply the framework of Section~\ref{sec:closure} to conformal risk control. Section~\ref{subsec:crc-setup} sets up the problem and two baseline procedures, and Section~\ref{subsec:orbit} characterizes validity exactly through the losses on hypothetical augmented samples. This characterization exposes the budget that a controller leaves unused, which recertification spends in Section~\ref{subsec:crc-refine}. Repeating the step leads to admissible controllers, which we illustrate on a small example (Section~\ref{subsec:iteration}). Section~\ref{subsec:extensions} extends the construction to multiple risks, nonconvex losses and test features, and discusses computation, and Section~\ref{sec:sim} reports a real-data study on hierarchical image classification.

	\subsection{Setup and two valid starting rules}\label{subsec:crc-setup}
	Conformal risk control \citep{angelopoulos2024crc} calibrates a family of actions so that a prescribed expected loss is controlled at a new test point. The actions are indexed by a parameter $\lambda$ in a finite ordered grid or a compact interval $\Lambda=[\lambda_-,\lambda_+]$, where larger values are more conservative. For example, $\lambda$ may be a threshold that determines a prediction set, with larger $\lambda$ giving larger sets. The quality of action $\lambda$ on an observation $z\in\Zc$ is measured by a known loss $\ell(z,\lambda)$, such as the indicator that the prediction set misses the label, or the fraction of true labels that it misses. Following the original work, we assume that
	\begin{equation}\label{eq:crc-assumptions}
		0\leq\ell(z,\lambda)\leq B,\qquad
		\ell(z,\lambda)\text{ is non-increasing and right-continuous in }\lambda,
		\qquad \ell(z,\lambda_+)\leq\alpha,
	\end{equation}
	where $0<\alpha<B$ is the target level. Thus the losses are bounded, a more conservative action never incurs a larger loss, and the most conservative action always meets the target.

	We observe calibration data $D=(Z_1,\ldots,Z_n)$ and must choose an action for a test observation $Z_{n+1}$, whose loss we cannot see. We assume only that $(Z_1,\ldots,Z_{n+1})$ is exchangeable; if the loss involves a trained prediction model, the model is held fixed, and exchangeability is assumed conditionally on its training data. The goal is a data-dependent action $A(D)\in\Lambda$ satisfying
	\[
	\E\,\ell(Z_{n+1},A(D))\leq\alpha
	\]
	under every finite exchangeable law on $\Zc^{n+1}$, and, subject to this constraint, as liberal as possible. We call such a rule $A$ a controller. The observation space and the loss family are fixed throughout, and controllers are always compared for the same observation space and loss family.

	Standard CRC bounds the unobserved test loss by $B$ and selects the smallest $\lambda$ for which the calibration losses, together with this bound, fit within the budget $(n+1)\alpha$:
	\begin{equation}\label{eq:crc}
		\CRC(D)=\min\left\{\lambda:\sum_{i=1}^n\ell(Z_i,\lambda)+B\leq (n+1)\alpha\right\}.
	\end{equation}
	The bound $B$ is the same for every $\lambda$, whereas the loss that the test point can incur at a given $\lambda$ may be much smaller, particularly for conservative actions. \citet{yeh2025training} therefore allow the bound to depend on $\lambda$. Given a known non-increasing, right-continuous envelope $U$ with
	\[
	\sup_{z\in\Zc}\ell(z,\lambda)\leq U(\lambda)\leq B,
	\]
	envelope CRC selects
	\begin{equation}\label{eq:env}
		\ENV(D)=\min\left\{\lambda:\sum_{i=1}^n\ell(Z_i,\lambda)+U(\lambda)\leq (n+1)\alpha\right\}.
	\end{equation}
	In both rules, the safe endpoint $\lambda_+$ is used if the displayed set is empty. Both rules are valid \citep{angelopoulos2024crc,yeh2025training}, and $U\leq B$ gives $\ENV\leq\CRC$ pointwise. The construction below improves any valid controller, envelope CRC included, and since the goal is an expected-loss bound, it fits the framework of Section~\ref{sec:closure}: the e-values that certify a controller show where it can be improved.

	\subsection{The exact risk budget on a hypothetical augmented sample}\label{subsec:orbit}
	Validity under exchangeability can be checked one augmented sample at a time. For an augmented multiset $S=(z_1,\ldots,z_{n+1})$ of $n+1$ points, let $S_{-i}$ denote $S$ with one occurrence of $z_i$ removed, and define
	\begin{equation}\label{eq:orbit-risk}
		r_i^A(S)=\ell(z_i,A(S_{-i})),\qquad
		T_A(S)=\sum_{i=1}^{n+1} r_i^A(S),\qquad
		\Delta_A(S)=(n+1)\alpha-T_A(S).
	\end{equation}
	Here $r_i^A(S)$ is the loss incurred when $z_i$ plays the role of the test point and the remaining $n$ points form the calibration data, $T_A(S)$ is the total of these held-out losses, and $\Delta_A(S)$ is the part of the budget $(n+1)\alpha$ that they leave unused. Repeated observations are counted with their multiplicities.

	\begin{lemma}[Exact leave-one-out characterization]\label{lem:orbit}
		A symmetric controller $A$ is valid under every finite exchangeable law if and only if
		\begin{equation}\label{eq:orbit-characterization}
			T_A(S)\leq (n+1)\alpha\qquad\text{for every deterministic augmented multiset }S.
		\end{equation}
	\end{lemma}
	\begin{proof}
		By symmetry of $A$ and exchangeability, the test loss has the same distribution as each held-out loss, so $\E\ell(Z_{n+1},A(Z_{1:n}))=(n+1)^{-1}\E T_A(Z_{1:n+1})$, which proves sufficiency. For necessity, fix $S$ and let the data be a uniformly random ordering of $S$. This law is exchangeable, and its risk is exactly $T_A(S)/(n+1)$.
	\end{proof}
	The lemma is closely connected to the orbit-average characterization of exchangeability e-values \citep{vovk2025eprediction,koning2026invariance}, and it is the risk-control counterpart of the universality of conformal prediction \citep[Proposition~2.9]{vovk2022algorithmic}. Its necessity part uses all finite exchangeable laws, which is why we require validity under exchangeability rather than under i.i.d. sampling alone (Appendix~\ref{app:iid}). %

	For a valid controller, $\Delta_A(S)\geq0$ is an exact account of the unused budget of the augmented sample $S$: it is the total amount by which the $n+1$ held-out losses on $S$ could increase without violating \eqref{eq:orbit-characterization}. It can be computed from the controller and the known loss once every point of $S$ is specified (see the example in Section~\ref{subsec:iteration}).
	Finally, we remark that the original CRC also connects to this view: its validity can be derived from conformal e-values \citep{vovk2025eprediction,koning2026invariance} that are allowed to depend on the action being certified (Appendix~\ref{app:crc-baseline}).

	\subsection{A baseline-preserving refinement}\label{subsec:crc-refine}
	Starting from any valid symmetric controller $A$, recertification spends $\Delta_A(S)$ by distributing it among the held-out positions. For example, with equal sharing,
	\begin{equation}\label{eq:equal-budget}
		b_i^A(S)=r_i^A(S)+\frac{\Delta_A(S)}{n+1},
		\qquad \sum_{i=1}^{n+1} b_i^A(S)=(n+1)\alpha.
	\end{equation}
	The new budgets still sum to $(n+1)\alpha$ on every augmented sample, and each exceeds the corresponding held-out loss of $A$ by $\Delta_A(S)/(n+1)$. The recertified controller selects the smallest action whose loss, for every possible test point $z$, stays within its new budget:
	\begin{equation}\label{eq:recursion}
		\quad(\rec A)(D)=\min\left\{\lambda\leq A(D):
		\ell(z,\lambda)-\ell(z,A(D))\leq\frac{\Delta_A(D\cup\{z\})}{n+1}\ \forall z\in\Zc\right\}.\quad
	\end{equation}
	The baseline action $A(D)$ satisfies every constraint, so the feasible set is nonempty, and on a finite grid, or on a compact interval under \eqref{eq:crc-assumptions}, its minimum is attained. We assume that the minimizer is implemented measurably. More generally, any feasible action no larger than $A(D)$ enjoys the following guarantee, so exact minimization is optional.

	\begin{theorem}[Uniform improvement]\label{thm:crc-improvement}
		For every valid symmetric baseline $A$, the controller $\rec A$ is valid and satisfies $\rec A\leq A$ pointwise. In particular, with $A^{(0)}=\ENV$ and $A^{(k+1)}=\rec A^{(k)}$,
		\begin{equation}\label{eq:improvement-chain}
			A^{(k+1)}(D)\leq A^{(k)}(D)\leq\ENV(D)\leq\CRC(D)
			\quad\text{for every }D,k,
		\end{equation}
		and every iterate controls the expected loss at level $\alpha$.
	\end{theorem}
	\begin{proof}
		Fix $S$. The constraint in \eqref{eq:recursion} with $D=S_{-i}$ and $z=z_i$ shows that each held-out loss of $\rec A$ exceeds that of $A$ by at most $\Delta_A(S)/(n+1)$. Summing over $i$ gives
		\[
		\sum_{i=1}^{n+1}\ell(z_i,(\rec A)(S_{-i}))
		\leq T_A(S)+\Delta_A(S)=(n+1)\alpha,
		\]
		so $\rec A$ is valid by Lemma~\ref{lem:orbit}. Since $A(D)$ is feasible in \eqref{eq:recursion}, $\rec A\leq A$, and the chain \eqref{eq:improvement-chain} follows by induction from $\ENV\leq\CRC$.
	\end{proof}

	\paragraph{Recertified CRC as an instance of generalized e-closure.}
	The improved controller $\rec A$ is obtained exactly by applying Proposition~\ref{prop:conditional} and Theorem~\ref{thm:closure} to the CRC problem, with unordered augmented samples as states. Let the data be the calibration multiset $D$, and let a state be an augmented multiset $S=(z_1,\ldots,z_{n+1})$. The class $\Pc_S$ consists of a single law, under which $D=S_{-I}$ for a position $I$ drawn uniformly from $\{1,\ldots,n+1\}$. Conditioning a finite exchangeable law on the multiset of its $n+1$ values yields such a law, and each such law arises from the exchangeable law that orders $S$ uniformly at random. When $D\subseteq S$, the held-out point $S\setminus D$ is determined by $S$ and $D$, and the loss is
	\[
	L(S,\lambda;D)=\ell(S\setminus D,\lambda),
	\]
	with $L(S,\lambda;D)=0$ when $D$ cannot arise from $S$. Validity in state $S$ is then exactly the constraint \eqref{eq:orbit-characterization} for that $S$, so validity in every state is equivalent to validity under every finite exchangeable law.

	The canonical e-collection of a valid controller $A$ is $E^A_S(D)=L(S,A(D);D)/\alpha$, whose expectation under $\Pc_S$ is $T_A(S)/((n+1)\alpha)$. Its shortfall from one is $\Delta_A(S)/((n+1)\alpha)$, the unused budget of \eqref{eq:orbit-risk}. Since each state specifies a single known law, recertification fills this shortfall without conditioning. In Proposition~\ref{prop:conditional} we take $V_S$ constant, the baseline envelope $g_S(D)=L(S,A(D);D)$, the bounds $H_S=T_A(S)/(n+1)$ and $B_S=\alpha$, and the increment $W_S(D)=\Delta_A(S)/(n+1)$ for $D\subseteq S$ (and zero otherwise). The recertified e-collection is
	\[
	E'_S(D)=\frac{g_S(D)+W_S(D)}{\alpha}=\frac{b_i^A(S)}{\alpha}\qquad\text{for }D=S_{-i},
	\]
	which has expectation exactly one by \eqref{eq:equal-budget}. At an observed $D$, only the states $S=D\cup\{z\}$ with $z\in\Zc$ impose constraints, and the certified menu \eqref{eq:closure} becomes
	\[
	\Cc_\alpha(E';D)=\left\{\lambda\in\Lambda:\ \ell(z,\lambda)-\ell(z,A(D))\leq\frac{\Delta_A(D\cup\{z\})}{n+1}\ \ \forall z\in\Zc\right\}.
	\]

	Each constraint bounds a non-increasing function of $\lambda$, and $A(D)$ satisfies all of them, so the menu is an upper set of $\Lambda$ that contains $A(D)$, and its smallest element is $(\rec A)(D)$. The restriction $\lambda\leq A(D)$ in \eqref{eq:recursion} is thus automatic: the recertified controller selects the least conservative certified action, and by Theorem~\ref{thm:closure} any other selection from the menu would be valid as well.

	\begin{remark}\label{rem:full-sample}
		Alternatively, one can view the augmented sample as a conditioning statistic rather than as the state. Take the full sample $X=(Z_1,\ldots,Z_{n+1})$ as data, of which only $D=(Z_1,\ldots,Z_n)$ is observed, and use a single state whose class contains all finite exchangeable laws, with loss $\ell(Z_{n+1},\lambda)$. This class is composite, but conditioning on the multiset $V$ of the values $Z_1,\ldots,Z_{n+1}$ removes the dependence on the law. Proposition~\ref{prop:conditional} then applies with $H(V)=T_A(V)/(n+1)$, $B(V)=\alpha$ and $W(X)=\Delta_A(V)/(n+1)$. The resulting certified menu depends on the unobserved test point, and requiring the action to be certified for every possible value of $Z_{n+1}$ recovers \eqref{eq:recursion}.
	\end{remark}

	\begin{algorithm}[t]
		\fontsize{10}{12}\selectfont
		\caption{One recertification of a known-loss CRC controller}\label{alg:recycle}
		\begin{algorithmic}[1]
			\REQUIRE A valid symmetric controller $A$, loss $\ell$, calibration multiset $D$, and level $\alpha$.
			\STATE Compute $a_0=A(D)$.
			\FOR{each hypothetical $z\in\Zc$}
			\STATE Compute $T_A(D\cup\{z\})=\ell(z,a_0)+\sum_{j=1}^n\ell(Z_j,A(D_{-j}\cup\{z\}))$.
			\STATE Set $s_A(D,z)=\alpha-T_A(D\cup\{z\})/(n+1)$.
			\ENDFOR
			\STATE Return the smallest $\lambda\leq a_0$ such that $\ell(z,\lambda)-\ell(z,a_0)\leq s_A(D,z)$ for every $z$.
		\end{algorithmic}
	\end{algorithm}

	Algorithm~\ref{alg:recycle} summarizes one round of recertification at an observed calibration multiset $D$. Each round evaluates the current controller on all replacement samples $D_{-j}\cup\{z\}$. Equal sharing is the simplest allocation of the unused budget. Other permutation-respecting allocations are also valid, and every valid pointwise improvement of a controller with monotone losses arises from some allocation (Appendix~\ref{app:allocations}); which allocation works best in one step depends on the problem.

	\subsection{Iteration to admissibility under convexity}\label{subsec:iteration}
	Since recertification preserves validity, it can be repeated, and each round can only improve the selected action. For convex losses, the iterates converge to an admissible controller that cannot be improved further, in the following pointwise sense.

	\begin{definition}[Pointwise admissibility for CRC]\label{def:crc-admissibility}
		A valid symmetric deterministic controller $A$ is pointwise admissible if no other measurable symmetric deterministic controller $C$ is valid under every finite exchangeable law and satisfies $C(D)\leq A(D)$ for every $D$, strictly at some $D$. The observation domain, loss family, sample size and action space remain fixed.
	\end{definition}
	Admissibility is thus a pointwise notion: controllers are compared on every calibration sample. This is meaningful because Lemma~\ref{lem:orbit} imposes a constraint on every augmented multiset, with no exceptional null sets, so changing a controller at even a single calibration multiset can affect its validity. The notion concerns only how conservative the selected actions are. An admissible controller need not be optimal for a particular utility, and several admissible controllers may coexist, as the example at the end of this section shows.

	\begin{theorem}[An admissible limit]\label{thm:crc-limit}
		Suppose $\Lambda=[\lambda_-,\lambda_+]$ and every $\ell(z,\cdot)$ in \eqref{eq:crc-assumptions} is convex. Suppose the exact inversions in \eqref{eq:recursion} are measurably implemented. From any valid baseline, the iterates decrease to a valid pointwise-admissible controller $A^{(\infty)}$. With $\rho=n/(n+1)$,
		\begin{equation}\label{eq:crc-rate}
			0\leq A^{(k)}(D)-A^{(\infty)}(D)
			\leq\rho^k (A^{(0)}(D)-A^{(\infty)}(D))
			\leq\rho^k(\lambda_+-\lambda_-).
		\end{equation}
		A valid controller is admissible if and only if it is an exact fixed point of $\rec$ in this setting.
	\end{theorem}
	The proof rests on a one-step comparison. Suppose that a valid controller $C$ satisfies $C\leq A$. On every augmented sample, each held-out loss of $C$ exceeds that of $A$ by a nonnegative amount, and by Lemma~\ref{lem:orbit} these increases sum to at most $\Delta_A(S)$, so each is at most $\Delta_A(S)$. By convexity, the action $\rho A(D)+C(D)/(n+1)$ increases each loss by at most a fraction $1/(n+1)$ of this amount, so it is feasible in \eqref{eq:recursion}, and therefore
	\begin{equation}\label{eq:key-comparison}
		\rec A\leq\rho A+C/(n+1).
	\end{equation}
	The iterates decrease to a limit, which is valid by the right continuity of the losses. Applying \eqref{eq:key-comparison} along the iterates and passing to the limit shows that no valid controller strictly dominates the limit, and taking $C=A^{(\infty)}$ gives the rate \eqref{eq:crc-rate}. Appendix~\ref{app:limit} gives the complete proof.

	We note that convexity is used only for the admissibility of the limit, and every finite round of recertification is valid without it. The limit theorem also uses an interval of actions, since on a grid the interpolated action in \eqref{eq:key-comparison} may be unavailable; Appendix~\ref{app:toy} gives an example. Many common prediction losses, such as miscoverage indicators, are step functions of $\lambda$. For such losses the iteration still improves the baseline validly, and Section~\ref{subsec:extensions} describes a different completion that reaches an admissible controller using only monotonicity.

	\paragraph{A two-type example.}
	A small example makes the iteration concrete. Let the observation space consist of two points, $\Zc=\{H,L\}$, with a high and a low loss,
	\begin{equation}\label{eq:toy-loss}
		\ell(H,\lambda)=1-\lambda,\qquad
		\ell(L,\lambda)=\tfrac12(1-\lambda),\qquad \lambda\in\Lambda=[0,1],
	\end{equation}
	and take $n=2$ and $\alpha=1/2$. The probabilities of the two points, and the dependence between observations, are arbitrary apart from exchangeability. We write multisets of observations as strings: $HL$ is the calibration sample with one point of each kind, and $HLL$ is an augmented sample, that is, a state in the formulation of Section~\ref{subsec:crc-refine}. The sharp envelope is $U(\lambda)=1-\lambda$, and Table~\ref{tab:toy} lists each controller on the three possible calibration samples.

	\begin{table}[hb]
		\centering\fontsize{10}{12}\selectfont
		\caption{The two-type example with $n=2$ and $\alpha=1/2$. Rows are the possible calibration samples and columns are controllers; entries are the selected actions, and smaller is preferred. The iterates start at envelope CRC. Every comparison holds for all exchangeable distributions on $\{H,L\}^3$.}\label{tab:toy}
		\begin{tabular}{@{}lccccc@{}}\toprule
			Calibration & CRC & Envelope $A^{(0)}$ & $A^{(1)}$ & $A^{(2)}$ & $A^{(\infty)}$\\\midrule
			$HH$ & $3/4$ & $1/2$ & $1/2$ & $1/2$ & $1/2$\\
			$HL$ & $2/3$ & $2/5$ & $23/60$ & $17/45$ & $3/8$\\
			$LL$ & $1/2$ & $1/4$ & $1/5$ & $31/180$ & $1/8$\\\bottomrule
		\end{tabular}
	\end{table}

	For instance, at $D=LL$ the envelope action is $1/4$. Appending $H$ gives the augmented sample $HLL$, whose held-out losses $3/4$, $3/10$ and $3/10$ total $27/20$. The equal allowance is $1/2-(27/20)/3=1/20$, so the constraint for $z=H$ reads $1/4-\lambda\leq1/20$, that is, $\lambda\geq1/5$, while appending $L$ permits every $\lambda\geq0$. Thus $A^{(1)}(LL)=1/5$. The entries converge geometrically, as Theorem~\ref{thm:crc-limit} predicts with $\rho=2/3$. At the limit, the validity constraints for $HHH$, $HHL$ and $HLL$ bind, which makes the limit admissible, whereas the constraint for $LLL$ is slack: an admissible controller may leave budget unused on some augmented samples. Appendix~\ref{app:toy} gives the exact iteration formulas.
	Admissible controllers need not be unique either. The controller with entries $(1/2,2/5,1/10)$ on $(HH,HL,LL)$ is also admissible and is incomparable with the limit, and the pointwise minimum of the two is invalid. Which admissible controller is preferable thus generally depends on the specific problem setting.

	\subsection{Extensions and computation}\label{subsec:extensions}\label{subsec:multi-crc}\label{subsec:practical}

	We now extend the construction to multiple risks, to nonconvex losses and to controllers that use test features, and finally discuss its computational cost.

	\paragraph{Multiple risks.}
	In many applications a single action must control multiple losses at once. A prediction set, for example, may be required to control both its miscoverage and the fraction of true labels it misses, or a whole family of losses indexed by a continuous severity level. CRC handles such families when the losses share the same action \citep[Section~4.3]{angelopoulos2024crc}, and recertification extends in the same way.

	Let $\ell_t(z,\lambda)$, $t\in\mathcal T$, be a family of losses indexed by an arbitrary, possibly uncountable, set $\mathcal T$, each satisfying \eqref{eq:crc-assumptions} with its own level $\alpha_t>0$ and the same action space $\Lambda$. A controller $A$ is \emph{jointly valid} if $\E\,\ell_t(Z_{n+1},A(D))\leq\alpha_t$ for every $t\in\mathcal T$ under every finite exchangeable law. By Lemma~\ref{lem:orbit}, applied to each risk, this holds if and only if
	\[
	T_{A,t}(S)=\sum_{i=1}^{n+1}\ell_t(z_i,A(S_{-i}))\leq(n+1)\alpha_t
	\qquad\text{for every augmented multiset $S$ and every $t\in\mathcal T$}.
	\]
	Since all losses are non-increasing in $\lambda$, the most conservative of the separate CRC choices, that is, their supremum over $t$, is jointly valid whenever it is measurable, and the same holds for envelope CRC. For a jointly valid $A$, recertification shares each loss-wise unused budget equally among the held-out positions and imposes the resulting constraints for all risks simultaneously:
	\begin{equation}\label{eq:multi-recycle}
		\begin{aligned}
			\Delta_{A,t}(S)&=(n+1)\alpha_t-T_{A,t}(S),\\
			(\rec A)(D)&=\min\left\{\lambda\leq A(D):
			\ell_t(z,\lambda)-\ell_t(z,A(D))\leq\frac{\Delta_{A,t}(D\cup\{z\})}{n+1}\ \ \forall z\in\Zc,\ t\in\mathcal T\right\}.
		\end{aligned}
	\end{equation}

	\begin{proposition}[Multiple risks]\label{prop:multi-risk}
		For every jointly valid symmetric controller $A$, the controller $\rec A$ in \eqref{eq:multi-recycle} is jointly valid and satisfies $\rec A\leq A$. If, moreover, $\Lambda=[\lambda_-,\lambda_+]$ and every $\ell_t(z,\cdot)$ is convex, the conclusions of Theorem~\ref{thm:crc-limit} hold for the iterates of \eqref{eq:multi-recycle}, with the same rate $\rho=n/(n+1)$, regardless of the index set $\mathcal T$.
	\end{proposition}

	The proof of Theorem~\ref{thm:crc-improvement} simply applies to each risk separately; see Appendix~\ref{app:limit}.
	The number of risks thus enters neither the rate nor the admissibility conclusion. It matters only for attaining the minimum in \eqref{eq:multi-recycle} measurably. Appendix~\ref{app:two-risk} works out a simple two-risk example with a continuum of observations, in which recertification converges to the largest valid rule dominating the joint envelope baseline, a strict improvement at every interior point.

	\paragraph{Nonconvex losses.}
	Theorem~\ref{thm:crc-limit} relies on convexity, which fails for many prediction losses, such as miscoverage indicators. For general monotone losses, admissibility can instead be characterized one calibration multiset at a time. Let $\nu_D(z)$ denote the number of copies of $z$ in $D$. Lowering the controller only at $D$, from $A(D)$ to $a<A(D)$, changes the held-out losses only on the augmented samples $D\cup\{z\}$, in each of which $z$ is held out $\nu_D(z)+1$ times. By Lemma~\ref{lem:orbit}, the change preserves joint validity if and only if
	\begin{equation}\label{eq:coordinate-crc}
		(\nu_D(z)+1)\{\ell_t(z,a)-\ell_t(z,A(D))\}
		\leq\Delta_{A,t}(D\cup\{z\})\qquad\forall z\in\Zc,\ t\in\mathcal T.
	\end{equation}

	\begin{proposition}\label{prop:coordinate}
		A jointly valid symmetric controller $A$ is pointwise admissible if and only if \eqref{eq:coordinate-crc} fails for every calibration multiset $D$ and every $a<A(D)$. If the observation space $\Zc$ is finite, the following sweep produces a pointwise-admissible controller that is no larger than $A$: visit the calibration multisets once in a fixed order, and at each visit replace $A(D)$ by the smallest $a$ satisfying \eqref{eq:coordinate-crc}, holding all other entries fixed.
	\end{proposition}
	The sweep uses only monotonicity: lowering one entry can only shrink the allowances of the others, so an entry that cannot be lowered when it is visited cannot be lowered later (Appendix~\ref{app:coordinate}). Its cost is a single pass over the controller table, which has one entry per calibration multiset.

	\paragraph{Test features.}
	In prediction problems, the feature $x$ of the test point is observed and only its label $y$ is unknown, so the controller may depend on it, $A(D,x)$. Validity then means $\E\,\ell_t(Z_{n+1},A(D,X_{n+1}))\leq\alpha_t$, a guarantee that is marginal over test features and labels, as for CRC itself. On an augmented sample, the held-out loss becomes $\ell_t(z_i,A(S_{-i},x_i))$, and the recertified controller at the observed $(D,x)$ is
	\[
	(\rec A)(D,x)=\min\left\{\lambda\leq A(D,x):
	\ell_t((x,y),\lambda)-\ell_t((x,y),A(D,x))\leq\frac{\Delta_{A,t}(D\cup\{(x,y)\})}{n+1}\ \ \forall y,t\right\}.
	\]
	The constraints range only over the possible labels $y$, while each unused budget $\Delta_{A,t}$ evaluates the current controller at the feature of every held-out point. Lemma~\ref{lem:orbit} continues to hold with these held-out losses, since its proof uses only the symmetry of $A$ in $D$ and exchangeability. The constraint for the label $y$ at $(D,x)$ is the constraint for the held-out copy of $(x,y)$ in $D\cup\{(x,y)\}$, so the proofs of Theorems~\ref{thm:crc-improvement} and~\ref{thm:crc-limit} carry over with $D$ replaced by $(D,x)$.

	\paragraph{Computation.}
	The risk-control guarantees hold after any number of rounds, so the number of rounds is a computational choice, whose cost depends on how the controller depends on the data.

	A single round is usually inexpensive. With $K$ candidate outcomes, $(\rec A)(D)$ requires $A(D)$ and the totals $T_A(D\cup\{z\})$, that is, the controller evaluated on the $nK$ replacement samples $D_{-j}\cup\{z\}$, each a recalibration with the prediction model held fixed. For CRC and envelope CRC, these recalibrations share most of their work. A replacement changes a single term of the calibration loss $\sum_{i=1}^n\ell(Z_i,\lambda)$, so the sum is computed once and adjusted for each replacement, and each replacement action is the smallest $\lambda$ satisfying a constraint that is monotone in $\lambda$. Such an action is found by a binary search for general losses and directly for structured ones, such as piecewise-linear or step losses.

	Repeated rounds are more demanding, because the replacements compound. Round $k$ evaluates $A^{(k-1)}$ on replacement samples, each of which requires $A^{(k-2)}$ on samples with a second point replaced, and so on. Hence $A^{(k)}(D)$ depends on the baseline at every calibration multiset obtained from $D$ by replacing up to $k$ of its points with candidate outcomes. The cost of many rounds is governed by how many distinct multisets arise and whether they can be shared across calibration samples and test points, and both depend on how the baseline depends on the data.
	\begin{itemize}
		\item \emph{Finitely many types.} Suppose that the loss depends on an observation only through one of $K$ types and that the baseline depends on the calibration data only through the type counts. Every iterate then inherits this property, so all the multisets above belong to a single table indexed by count vectors. The table has $\binom{n+K-1}{K-1}$ entries, polynomially many in $n$ for fixed $K$, and each round is one pass over it. It does not depend on the observed data, so it is computed once and reused for every calibration sample and test point.
		\item \emph{Individual scores or features.} When the baseline depends on individual observations, for instance through continuous scores or test features, no such table exists. The replacement samples must then be tracked separately for each calibration sample and test point, and their number grows exponentially in the number of calibration points whose loss is not identically zero. Exact reductions, such as grouping points with identical scores, can still make many rounds feasible (Appendix~\ref{app:simulation-implementation}).
		\item \emph{Uncountable outcomes or risks.} Here even a single round requires care. The constraints in \eqref{eq:multi-recycle} must hold for a continuum of $z$ or $t$, and checking them requires a certified optimization or a uniform approximation bound, since finitely many sampled outcomes do not certify feasibility.
	\end{itemize}
	Two observations reduce the cost in every case. First, the iterates decrease and the losses are non-increasing, so stopping at a data-dependent round no later than a fixed maximum incurs at most the loss of the maximal round and preserves validity. Computation can therefore stop once a computable bound shows that further rounds cannot change the action appreciably. Second, replacing the allowances $\Delta_{A,t}$ by conservative lower bounds preserves validity and the comparison with the baseline, although the limit may then fall short of admissibility. Exact recertification thus trades computation for a uniform improvement.

	\subsection{Real-data study}\label{sec:sim}
	We evaluate recertification on hierarchical image classification, one of the applications of CRC in \citet{angelopoulos2024crc}, and compare it with ordinary CRC and with envelope CRC \citep{yeh2025training}.

	\subsubsection{Hierarchical classification}\label{sec:simulation-hierarchy}
	We consider the hierarchical image classification setting, where each observation is $Z_i=(X_i,Y_i)$. Here, each $X_i\in\mathcal X$ is an image and $Y_i\in\mathcal Y=\{1,\ldots,K_1\}$ is its fine-class label. The labels are partitioned into $K_2$ superclasses $G_1,\ldots,G_{K_2}$, which form a tree of depth two with a root, the superclasses and the fine-class leaves. We identify each node with the set of its descendant labels, so that the root is $\mathcal Y$.

	A fixed, pretrained classifier outputs class probabilities $p(x)=(p_1(x),\ldots,p_{K_1}(x))$. Let $f(x)=\arg\max_k p_k(x)$, with ties broken deterministically, and let $g(x)$ be the superclass containing $f(x)$. For an image $x$, predictions are restricted to the three nodes $\mathcal Y$, $g(x)$ and $\{f(x)\}$ on this path. For a reported node $v$ and a true class $y$, the hierarchical loss is
	\[
	L(v,y)=\frac12\min_{u\in\mathrm{anc}(y)}d(v,u),
	\]
	where $d$ is the tree distance and $\mathrm{anc}(y)$ consists of the leaf $\{y\}$ and its ancestors. Thus $L(v,y)=0$ when $y\in v$, and otherwise $L(v,y)$ is half the number of upward edges needed to reach a node containing $y$.

	To obtain a continuous action space, we allow randomization between adjacent depths. For $a\in[0,2]$, let $N_a(x)$ report the root, the superclass and the leaf with probabilities $((1-a)_+,\,1-|a-1|,\,(a-1)_+)$, where $r_+=\max\{r,0\}$. Its mean depth is $a$, and its expected loss is
	\begin{equation}
		\ell((x,y),a)=\E[L(N_a(x),y)]=
		\begin{cases}
			0, & y=f(x),\\
			(a-1)_+/2, & y\in g(x)\setminus\{f(x)\},\\
			a/2, & y\notin g(x).
		\end{cases}
		\label{eq:simulation-hierarchy-loss}
	\end{equation}
	Given calibration data $D=(Z_1,\ldots,Z_n)$ and a target $\alpha\in(0,1)$, a procedure chooses a depth rule $\widehat a:\mathcal X\to[0,2]$ and reports $N_{\widehat a(x)}(x)$ for a test image $x$. The aim is to make predictions as specific as possible subject to
	\[
	\E\,\ell\bigl(Z_{n+1},\widehat a(X_{n+1})\bigr)\leq\alpha,
	\]
	where $Z_{n+1}$ is a test observation exchangeable with $D$, the expectation is over the calibration and test data, and the randomization of the prediction is already integrated into $\ell$. This is the setting of Section~\ref{sec:crc} with test features, in which the depth rule is the controller.

	We consider two families of depth rules $a_t$, indexed by a parameter $t$. The first uses a common randomized depth, $a_t(x)=t$ for $t\in[0,2]$: every image receives the same distribution over depths, although the reported superclass and leaf still depend on the image through $g(x)$ and $f(x)$. The second adapts the depth to the confidence of the classifier. Write $p_f(x)=p_{f(x)}(x)$ for the probability of the most likely class and $p_g(x)=\sum_{k\in g(x)}p_k(x)$ for the total probability of its superclass. For a threshold $t\in[0,1]$, set
	\[
	a_t(x)=\ind\{t>1-p_g(x)\}+\ind\{t>1-p_f(x)\}.
	\]
	This rule reports the root when $p_g(x)\le1-t$, the superclass when $p_f(x)\le1-t<p_g(x)$, and the leaf when $p_f(x)>1-t$, so more confident images receive more specific predictions.

	In both families, larger values of $t$ are less conservative, so the parameter $\lambda$ of Section~\ref{sec:crc} corresponds to $-t$. CRC selects the largest $t$ satisfying
	\[
	\sum_{i=1}^n\ell(Z_i,a_t(X_i))+1\le(n+1)\alpha,
	\]
	where $1$ is the largest possible loss. Envelope CRC replaces this bound by the sharper envelope $U(t)$, which equals $t/2$ for common depth and $1$ for every $t>0$ for confidence-adaptive scores; in the latter family, envelope CRC therefore coincides with CRC. Starting from envelope CRC, we apply one round and $500$ rounds of recertification with equal sharing. Each round recomputes the unused budgets on the augmented samples and increases the depth as far as they permit. In the common-depth family, the recertified rule still assigns one randomized depth to all images and depends on the calibration data only through the counts of the three cases in \eqref{eq:simulation-hierarchy-loss}, so it is a finite table of the kind discussed in Section~\ref{subsec:extensions}. In the confidence-adaptive family, the recertified rule depends on the test image and may randomize between adjacent depths beyond the deterministic threshold prediction.

	\subsubsection{Results}\label{sec:simulation-results}
	We use CIFAR-100, which has $K_1=100$ fine classes and $K_2=20$ superclasses of five classes each, so the root, superclass and leaf predictions contain $100$, $5$ and $1$ labels. The classifier is a frozen ResNet-18, and the data pool consists of its cached predictions on the $10{,}000$ test images \citep{gonzalez2026embedded}. We report the mean hierarchical loss (risk), the mean node size $|v|$, and the correct specificity $\ind\{y\in v\}[1-\log|v|/\log K_1]$, which rewards predictions that are both correct and informative. All three metrics are averaged analytically over the randomization of the prediction, and gains are absolute differences in correct specificity, in percentage points.

	We fix $n=50$ and $\alpha\in\{0.02,0.05,0.10,0.15\}$, and average over $300$ random splits of the pool, each with $50$ calibration images and $16$ disjoint test images. The same splits are used for all families, methods and targets, and standard errors are computed across splits. Appendix~\ref{app:simulation-implementation} gives the implementation details.

	Table~\ref{tab:simulation-comparison} reports the results.
	All methods have empirical risk below the target in every setting. In the common-depth family, envelope CRC is much less conservative than CRC, particularly at small targets, and leaves little of the risk budget unused. Recertification therefore adds only a small gain on top of it, which grows with the target from none at $\alpha=0.02$ to $0.44$ percentage points at $\alpha=0.15$. In the confidence-adaptive family, envelope CRC coincides with CRC, and the entire improvement comes from recertification. At $\alpha=0.10$, $500$ rounds raise correct specificity from $67.57\%$ to $68.24\%$ and reduce the mean node size from $19.40$ to $17.03$ labels; at $\alpha=0.02$, the gain is $1.71$ percentage points.

\begin{figure}[!htbp]
		\centering
		\includegraphics[width=\textwidth]{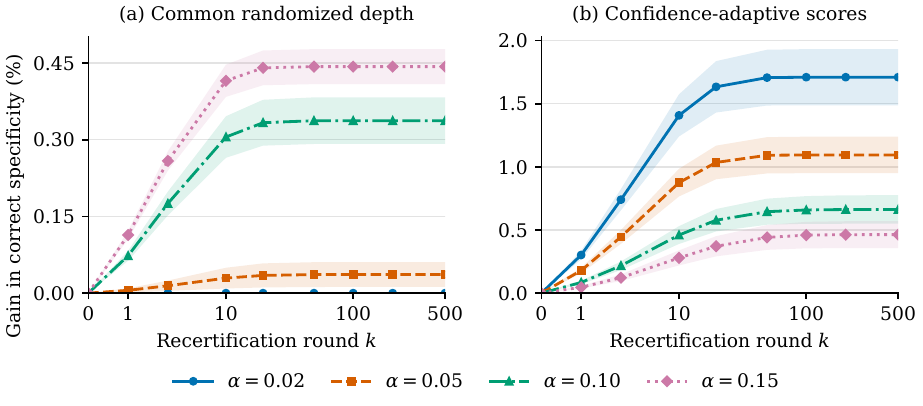}
		\caption{Gains in correct specificity of recertification beyond envelope CRC at $n=50$. Each panel shows mean paired gains over 300 splits, with shaded pointwise 95\% Student-$t$ intervals. The horizontal axis is logarithmic in $k+1$, where $k$ is the number of rounds.}
		\label{fig:simulation-trajectory}
	\end{figure}

	\begin{table}[!htbp]
		\centering\small
		\setlength{\tabcolsep}{3.4pt}
		\caption{Hierarchical classification on CIFAR-100 with $n=50$, averaged over 300 paired splits. One round and 500 rounds denote recertification of envelope CRC. CS is correct specificity (\%) and size is the mean number of labels in the prediction. Time is in milliseconds per build of the reusable common-depth controller and per test image for confidence-adaptive scores.}
		\label{tab:simulation-comparison}
		\begin{tabular}{rlrrrrrrrr}
			\toprule
			& & \multicolumn{4}{c}{Common randomized depth} & \multicolumn{4}{c}{Confidence-adaptive scores} \\
			\cmidrule(lr){3-6}\cmidrule(lr){7-10}
			$\alpha$ & Method & Risk & CS (\%) & Size & Time & Risk & CS (\%) & Size & Time \\
			\midrule
			0.02 & CRC & 0.00044 & 0.282 & 99.50 & 9.347 & 0.01219 & 41.903 & 55.06 & 0.003 \\
			& Envelope CRC & 0.01961 & 12.538 & 77.96 & 9.962 & 0.01219 & 41.903 & 55.06 & 0.003 \\
			& One round & 0.01961 & 12.538 & 77.96 & 9.907 & 0.01324 & 42.205 & 54.45 & 0.004 \\
			& 500 rounds & 0.01961 & 12.538 & 77.96 & 194.5 & 0.01814 & 43.613 & 51.58 & 2.885 \\
			\addlinespace
			0.05 & CRC & 0.03415 & 21.766 & 61.76 & 8.867 & 0.04052 & 60.032 & 32.97 & 0.003 \\
			& Envelope CRC & 0.04884 & 30.989 & 45.62 & 8.997 & 0.04052 & 60.032 & 32.97 & 0.003 \\
			& One round & 0.04887 & 30.994 & 45.62 & 9.341 & 0.04175 & 60.211 & 32.50 & 0.004 \\
			& 500 rounds & 0.04907 & 31.025 & 45.61 & 198.0 & 0.04803 & 61.127 & 30.06 & 5.127 \\
			\addlinespace
			0.10 & CRC & 0.08748 & 49.556 & 15.46 & 8.877 & 0.09052 & 67.572 & 19.40 & 0.003 \\
			& Envelope CRC & 0.09645 & 52.883 & 10.65 & 8.937 & 0.09052 & 67.572 & 19.40 & 0.003 \\
			& One round & 0.09700 & 52.956 & 10.64 & 9.275 & 0.09156 & 67.657 & 19.10 & 0.004 \\
			& 500 rounds & 0.09898 & 53.220 & 10.58 & 195.2 & 0.09880 & 68.235 & 17.03 & 28.2 \\
			\addlinespace
			0.15 & CRC & 0.13944 & 61.441 & 3.72 & 9.325 & 0.14208 & 70.550 & 10.77 & 0.003 \\
			& Envelope CRC & 0.14462 & 62.279 & 3.37 & 9.157 & 0.14208 & 70.550 & 10.77 & 0.003 \\
			& One round & 0.14543 & 62.393 & 3.35 & 9.755 & 0.14279 & 70.597 & 10.59 & 0.004 \\
			& 500 rounds & 0.14780 & 62.722 & 3.28 & 201.3 & 0.14955 & 71.014 & 8.92 & 138.2 \\
			\bottomrule
		\end{tabular}
		\par\smallskip
		\parbox{\textwidth}{\footnotesize Timings use an Intel Core i7-12700KF with Python/Numba and exclude model inference and compilation warm-up. Common-depth builds and the faster adaptive methods average 20 repetitions; 500-round adaptive timings cover all 4800 test images, including table construction.}
	\end{table}
	Figure~\ref{fig:simulation-trajectory} shows how the gains beyond envelope CRC accumulate over rounds. A single round recovers between $10\%$ and $26\%$ of the gain after $500$ rounds. At $\alpha=0.10$, for example, one round gains $0.073$ percentage points for common depth and $0.085$ for confidence-adaptive scores, compared with $0.337$ and $0.663$ after $500$ rounds. The first $50$ rounds recover at least $95\%$ of the final gain in every setting, and later rounds change little.

	The two families locate the conservativeness of CRC in different places. In the common-depth family, most of it lies in the bound on the unseen loss: the envelope $U(t)=t/2$ is much smaller than one at shallow depths, and envelope CRC removes most of the slack. In the confidence-adaptive family, the envelope equals the worst-case bound, yet recertification still produces more informative predictions, using budget that no choice of envelope can recover.

	The possible increase in attained risk is bounded by the gap between the target level and the baseline risk. Since the risk of a controller $A$ equals $\E\,T_A(S)/(n+1)$ for an augmented sample $S$, the gap between $\alpha$ and the risk is exactly the expected unused budget $\E\,\Delta_A(S)/(n+1)$, which is what recertification spends. In the common-depth family, envelope CRC leaves only about $2\%$ to $4\%$ of the target unused, and at $\alpha=0.02$ its risk is within sampling error of the target, so there is almost nothing to recover. In the confidence-adaptive family, CRC leaves between $5\%$ and $39\%$ unused, and the gains are correspondingly larger. In this family, the gains also shrink as the target grows, partly because more predictions are already as specific as possible: at $\alpha=0.15$, $86.3\%$ of the baseline predictions are leaves.

	The runtimes reflect the two regimes of Section~\ref{subsec:extensions}. The common-depth controller is a single table, built in about $0.2$ seconds for $500$ rounds and reused across splits and test images. Confidence-adaptive recertification is computed separately for each test image, and $500$ rounds take from about $3$ to $140$ milliseconds per image, with the cost growing in the target.

	\section{Improving randomized e-BH}\label{sec:rebh}\label{sec:applications}
	We now return to multiple testing, the setting of the e-closure principle that inspired this work \citep{xu2026closure}, and recertify a randomized version of the e-BH procedure. Let $E=(E_1,\ldots,E_K)$ be e-values for hypotheses $H_1,\ldots,H_K$, with arbitrary dependence among them. In the framework of Section~\ref{sec:closure}, the state is the unknown set $T\subseteq\{1,\ldots,K\}$ of true nulls, whose class $\Pc_T$ consists of all laws under which $\E_PE_j\leq1$ for every $j\in T$, and errors are measured by the false discovery proportion $\FDP_T(R)=|R\cap T|/(|R|\vee1)$ of a rejection set $R$, with budget $\alpha$. Section~\ref{subsec:rebh} recalls how e-BH fits this framework. Section~\ref{subsec:rebh-slack} casts randomized e-BH as an instance of generalized e-closure, with randomized decisions as actions, and identifies the budget it leaves unused. Section~\ref{subsec:rebh-recert} recertifies it with the e-collection of e-BH, Section~\ref{subsec:rebh-mean} with the larger e-collection of closed e-BH, and Section~\ref{subsec:rebh-sim} reports simulations.

	\subsection{e-BH and self-consistency}\label{subsec:rebh}
	For an observed vector $e$, let $e_{[1]}\geq\cdots\geq e_{[K]}$ be its ordered values, let $r(j)$ be the rank of $e_j$, and let $R_k(e)$ be the set of the $k$ hypotheses with the largest e-values, with ties broken by a fixed convention. The e-BH procedure of \citet{wangramdas2022} rejects $R_k(e)$ for the largest $k$ with $k\,e_{[k]}\geq K/\alpha$. With rejection sets as actions, its validity is explained by the e-collection
	\begin{equation}\label{eq:ebh-collection}
		E^{\mathrm{eBH}}_T=\frac1K\sum_{j\in T}E_j,
	\end{equation}
	whose expectation under $\Pc_T$ is at most $|T|/K\leq1$. Certification generally imposes one constraint for each state, here one for each of the $2^K$ sets of true nulls, but for this e-collection the constraints collapse to $K$ simple ones. Since both sides of $\FDP_T(R)\leq\alpha E^{\mathrm{eBH}}_T$ are sums over the hypotheses in $T$, the condition holds for every $T$ if and only if it holds for every singleton $T=\{j\}$, that is, if and only if $e_j\geq K/(\alpha|R|)$ for every $j\in R$. This is the self-consistency condition of \citet{blanchard2008two}, which \citet{wangramdas2022} used to establish the validity of e-BH under arbitrary dependence. The certified sets are thus exactly the self-consistent ones, and e-BH is the largest of them (Appendix~\ref{app:rebh-menu}).

	\subsection{Randomized e-BH as generalized e-closure}\label{subsec:rebh-slack}
	The e-BH procedure can be conservative, because an e-value contributes only once it crosses one of the thresholds $K/(\alpha k)$, and an e-value just below its threshold is wasted. \citet{xu2023randomization} showed that external randomization recovers part of this loss. The most powerful variant in their simulations, U-eBH, draws a single $U\sim\mathrm{Unif}(0,1)$ independent of $E$, boosts the e-values to $(E_1/U,\ldots,E_K/U)$, and applies e-BH to the boosted values; that is, it rejects $R_{\kappa_0}(E)$ with
	\[
	\kappa_0(e,U)=\max\bigl(\{k:U\leq\alpha k\,e_{[k]}/K\}\cup\{0\}\bigr).
	\]
	U-eBH rejects at least the e-BH set for every value of $U$ and controls the FDR at level $\alpha$. Its validity is not a consequence of applying e-BH to e-values, since $E_j/U$ has infinite expectation whenever $\Prob(E_j>0)>0$, but it becomes transparent once the randomization is made part of the action. A second variant, J-eBH, divides each e-value by its own uniform random variable drawn independently of the others and of $E$, and applies e-BH to the boosted values; it is also valid (Appendix~\ref{app:rebh-general}).

	\paragraph{Actions and losses.}
	Like e-BH, U-eBH rejects the hypotheses with the largest e-values, and only the number of rejections is random. We therefore consider \emph{monotone} rules, which reject $R_\kappa(e)$ for a random number $\kappa$ of rejections, and take an action to be a distribution $p=(p_0,\ldots,p_K)$ of $\kappa$. A \emph{boosting rule} chooses such a distribution $p(e)$ after observing the e-values, draws $\kappa\sim p(e)$ independently of the data, and rejects $R_\kappa(e)$. The loss of the action $p$ in state $T$ is the false discovery proportion averaged over this draw,
	\begin{equation}\label{eq:rebh-loss}
		\bar L(T,p;e)=\sum_{k=1}^Kp_k\,\frac{|T\cap R_k(e)|}{k}.
	\end{equation}
	This loss depends on the observation through the ranking of the e-values, which the framework of Section~\ref{sec:closure} allows, and a boosting rule is valid in the sense of \eqref{eq:risk-target} exactly when the procedure it implements controls the FDR at level $\alpha$. Randomization can thus also be viewed as part of the action, with probability distributions as the actions, as described in Section~\ref{subsec:menus}. U-eBH is the boosting rule whose distribution $p^0(e)$ has the tails
	\begin{equation}\label{eq:uebh-tails}
		\Prob(\kappa_0\geq r\mid E=e)=\min\left\{1,\frac\alpha K\max_{k\geq r}k\,e_{[k]}\right\},\qquad r=1,\ldots,K.
	\end{equation}
	Indeed, dividing every e-value by the same $U$ preserves their ranking, so e-BH applied to the boosted values rejects $k$ hypotheses when $k$ is the largest number satisfying $k\,e_{[k]}/U\geq K/\alpha$, that is, $U\leq\alpha k\,e_{[k]}/K$. Hence $\kappa_0\geq r$ exactly when $U\leq\alpha k\,e_{[k]}/K$ for some $k\geq r$, equivalently when $U\leq(\alpha/K)\max_{k\geq r}k\,e_{[k]}$, and because $U$ is uniform on $(0,1)$ and independent of $E$, this event has the conditional probability \eqref{eq:uebh-tails}.

	\paragraph{The certified menu.}
	The hypothesis of rank $r$ is rejected exactly when $\kappa\geq r$, and it then receives the share $1/\kappa$ of the false discovery proportion, if it is null. The loss \eqref{eq:rebh-loss} therefore splits over the hypotheses in $T$,
	\[
	\bar L(T,p;e)=\sum_{j\in T}c_{[r(j)]}(p),\qquad c_{[r]}(p)=\sum_{k\geq r}\frac{p_k}{k},
	\]
	where $c_{[r]}(p)$ is the expected share charged to the hypothesis of rank $r$, should it be null. Both sides of $\bar L(T,p;e)\leq\alpha E^{\mathrm{eBH}}_T(e)$ are again sums over the hypotheses in $T$, so the reduction of Section~\ref{subsec:rebh} applies verbatim, and the certified menu of $E^{\mathrm{eBH}}$ at $e$ is
	\begin{equation}\label{eq:rebh-budget}
		\Cc_\alpha(E^{\mathrm{eBH}};e)=\Bigl\{p:\ c_{[r]}(p)\leq\frac\alpha K\,e_{[r]}\ \text{ for } r=1,\ldots,K\Bigr\}.
	\end{equation}
	Each hypothesis may thus be charged at most its own share of the budget, a self-consistency condition that need only hold on average over the randomization. The menu is a polytope of distributions, and by Theorem~\ref{thm:closure} every boosting rule that selects from it has FDR at most $\alpha|T|/K\leq\alpha$.

	U-eBH selects from the menu \eqref{eq:rebh-budget}, which proves its validity, but its distribution usually satisfies the constraints with strict inequality. An intuitive explanation is that the common boosting factor $U$ does not respect the different slack structures. Indeed, dividing every e-value by the same $U$ ties the hypotheses together: U-eBH rejects at least $r$ hypotheses when e-BH, applied to the boosted values, selects some set $R_k(e)$ with $k\geq r$, which happens when $U\leq\alpha k\,e_{[k]}/K$ and therefore depends only on the smallest e-value $e_{[k]}$ in that set. As a result, a hypothesis can be charged as if its e-value were that of a lower-ranked hypothesis, and part of its budget then goes unused (Appendix~\ref{app:rebh-menu}).

	\subsection{Recertifying U-eBH}\label{subsec:rebh-recert}
	Recertification spends this unused budget. Like every valid rule, U-eBH is certified by its canonical e-collection, its averaged loss divided by the budget, $\bar L(T,p^0(e);e)/\alpha$. As shown above, this e-collection lies below $E^{\mathrm{eBH}}$, hypothesis by hypothesis, and the gap is exactly the unused budget. Recertification replaces the canonical e-collection by the larger e-collection $E^{\mathrm{eBH}}$. This step involves no unknown conditional expectation: the input e-values supply the budget directly, and the averaged losses are known because the law of the external randomization is known (Appendix~\ref{app:rebh-menu} spells out the correspondence with Proposition~\ref{prop:conditional}). The certified menu of the recertified e-collection is \eqref{eq:rebh-budget}. It contains U-eBH, but also distributions that do not arise from boosting all e-values by a common factor, and any of them may be selected. We select the one that rejects the most.

	To find this distribution, we turn the budget constraints in \eqref{eq:rebh-budget}, which bound the charges of individual hypotheses, into bounds on the quantities that measure power: the probabilities $\Prob(\kappa\geq r)$ of rejecting at least $r$ hypotheses. Recall that the hypothesis of rank $j$ is rejected exactly when $\kappa\geq j$, and then receives the share $1/\kappa$, so its charge is $c_{[j]}(p)=\E[\ind\{\kappa\geq j\}/\kappa]$. On an event $\{\kappa=k\}$ with $k\geq r$, the shares of the $k$ rejected hypotheses sum to one: the top $r$ hypotheses receive $r/k$ together, and each hypothesis of rank $j\in\{r+1,\ldots,k\}$ receives $1/k$. Averaging over the events $\{\kappa=k\}$ with $k\geq r$, the first part contributes $r\,\E[\ind\{\kappa\geq r\}/\kappa]=r\,c_{[r]}(p)$, and the second contributes $\sum_{j>r}\E[\ind\{\kappa\geq j\}/\kappa]=\sum_{j>r}c_{[j]}(p)$, because a hypothesis of rank $j>r$ is rejected only on these events. Hence
	\[
	\Prob(\kappa\geq r)=r\,c_{[r]}(p)+\sum_{j>r}c_{[j]}(p).
	\]
	Bounding each charge by its budget now bounds the probability of rejecting at least $r$ hypotheses directly. The top $r$ hypotheses can contribute at most $r$ times the budget of the hypothesis of rank $r$, and each lower-ranked hypothesis at most its own budget. Every distribution in the certified menu \eqref{eq:rebh-budget} therefore obeys
	\begin{equation}\label{eq:rebh-tails}
		\Prob(\kappa\geq r\mid E=e)\leq q^*_r(e)=\min\left\{1,\frac\alpha K\Bigl(r\,e_{[r]}+\sum_{j>r}e_{[j]}\Bigr)\right\},\qquad r=1,\ldots,K.
	\end{equation}
	These bounds depend on the e-values alone, and a single distribution attains all $K$ of them at once; when $(\alpha/K)\sum_je_j\leq1$, it makes every constraint in \eqref{eq:rebh-budget} tight (Appendix~\ref{app:rebh-proof}). The recertified rule selects this distribution and implements it with the seed of U-eBH, as summarized in Algorithm~\ref{alg:rebh}. Computation requires only sorting the e-values and forming partial sums. U-eBH also selects from the menu, so its tails \eqref{eq:uebh-tails} lie below $q^*_r$ -- they value the $k$ rejected hypotheses as if each had the smallest e-value among them.

	\begin{algorithm}[t]
		\fontsize{10}{12}\selectfont
		\caption{Recertified U-eBH}\label{alg:rebh}
		\begin{algorithmic}[1]
			\REQUIRE E-values $e_1,\ldots,e_K$, level $\alpha$, and $U\sim\mathrm{Unif}(0,1)$ drawn independently of the e-values.
			\STATE Order the e-values as $e_{[1]}\geq\cdots\geq e_{[K]}$, with ties broken by a fixed convention.
			\STATE Compute the partial sums $s_r=\sum_{j=r}^Ke_{[j]}$ for $r=1,\ldots,K$, and set $s_{K+1}=0$.
			\FOR{$r=1,\ldots,K$}
			\STATE Set $q^*_r=\min\{1,(\alpha/K)(r\,e_{[r]}+s_{r+1})\}$.
			\ENDFOR
			\STATE Set $\kappa^*=\max(\{r:U\leq q^*_r\}\cup\{0\})$.
			\ENSURE The rejection set $R_{\kappa^*}(e)$, consisting of the $\kappa^*$ hypotheses with the largest e-values.
		\end{algorithmic}
	\end{algorithm}

	The following proposition collects the properties of the recertified rule, which we regard as recertified randomized e-BH or, more precisely, recertified U-eBH (RU-eBH).

	\begin{proposition}[Recertified U-eBH]\label{prop:rebh}
		RU-eBH has the following properties.
		\begin{enumerate}[label=(\roman*),leftmargin=*,itemsep=0pt]
			\item Its FDR is at most $\alpha|T|/K\leq\alpha$ in every state $T$ and under every law in $\Pc_T$.
			\item It contains the U-eBH rejection set, and hence the e-BH rejection set, for every $e$ and $U$.
			\item Among boosting rules that select from \eqref{eq:rebh-budget}, it has the stochastically largest number of rejections: $\Prob(\kappa\geq r\mid E=e)\leq\Prob(\kappa^*\geq r\mid E=e)$ for every $r$ and $e$.
		\end{enumerate}
	\end{proposition}
	In particular, the probability of rejecting at least one hypothesis rises from $(\alpha/K)\max_kk\,e_{[k]}$ to $(\alpha/K)\sum_je_j$, capped at one. RU-eBH draws on the budget of every hypothesis, whereas U-eBH draws on that of a single rejection set.
	We also note that restricting attention to monotone rules costs nothing in the expected number of rejections: for every $e$, no randomized rule over arbitrary rejection sets that is certified by $E^{\mathrm{eBH}}$ rejects more hypotheses in expectation than RU-eBH, although such rules need not be stochastically smaller (Proposition~\ref{prop:rebh-general} in Appendix~\ref{app:rebh-general}). Finally, RU-eBH can be written as e-BH applied to the randomized e-values $(K/\alpha)\ind\{j\in R_{\kappa^*}\}/(|R_{\kappa^*}|\vee1)$ (Corollary~\ref{cor:rebh-evalues}).

	\subsection{Recertifying closed e-BH}\label{subsec:rebh-mean}
	The optimality in Proposition~\ref{prop:rebh} is relative to the e-collection $E^{\mathrm{eBH}}$, and a larger e-collection permits a further improvement. A simple example exploits the fact that the bound $\alpha|T|/K$ in Proposition~\ref{prop:rebh}(i) is strictly below $\alpha$ unless every hypothesis is null. The minimally adaptive principle \citep{solari2017minimally,ignatiadis2024evalues} uses this slack: it runs a procedure at the inflated level $K\alpha/(K-1)$, but only when the procedure at level $\alpha$ rejects at least one hypothesis, which keeps the FDR under the global null at most $\alpha$. In our framework, this corresponds to replacing $1/K$ by $1/(K-1)$ in \eqref{eq:ebh-collection} for every state other than the full set, which still gives an e-collection because such a state contains at most $K-1$ true nulls.
	Applying recertification to this collection would therefore yield a more powerful rule, and, as a result, RU-eBH is not admissible among all FDR procedures (Appendix~\ref{app:rebh-ma}).

	A much larger e-collection is introduced by \citet{xu2026closure},
	\begin{equation}\label{eq:mean-collection}
		E^{\overline{\mathrm{eBH}}}_T=\frac1{|T|}\sum_{j\in T}E_j\qquad(T\neq\varnothing),
	\end{equation}
	whose expectation under $\Pc_T$ is at most one and which certifies the rejection set of closed e-BH. It normalizes by the number of true nulls rather than by $K$, and thus adapts to the unknown proportion of true nulls.

	For a monotone rule, the averaged false discovery proportion in state $T$ is still the sum $\sum_{j\in T}c_{[r(j)]}(p)$ of the charges, but the budget $\alpha E^{\overline{\mathrm{eBH}}}_T$ no longer splits into budgets for individual hypotheses. The certified menu of $E^{\overline{\mathrm{eBH}}}$ at $e$ therefore consists of the distributions $p$ that satisfy all $2^K-1$ constraints
	\begin{equation}\label{eq:mean-menu}
		\sum_{j\in T}c_{[r(j)]}(p)\leq\frac{\alpha}{|T|}\sum_{j\in T}e_j\qquad(T\neq\varnothing),
	\end{equation}
	which do not reduce to $K$ separate ones.

	Nevertheless, the menu has enough structure that the rule with the largest expected number of rejections can be found in closed form.
	The key is to describe a monotone rule by its cumulative charges from the bottom. Let $S_s(p)=\sum_{r\geq s}c_{[r]}(p)$ be the total charge of the $K-s+1$ hypotheses with the smallest e-values. A hypothesis is rejected whenever a lower-ranked one is, with the same share, so the charges are non-increasing in the rank and $s\mapsto S_s(p)$ is convex. Conversely, every non-increasing convex sequence with $S_{K+1}=0$ and $S_1\leq1$ arises from a unique distribution $p$. The constraint \eqref{eq:mean-menu} for the set $\{s,\ldots,K\}$ caps $S_s(p)$ at $\alpha m_s$, where $m_s=(K-s+1)^{-1}\sum_{j\geq s}e_{[j]}$ is the mean e-value of that set, and $S_1(p)=\Prob(\kappa\geq1)$ is also at most one. Let $\beta_1=\min\{1,\alpha m_1\}$, $\beta_s=\alpha m_s$ for $2\leq s\leq K$ and $\beta_{K+1}=0$. These caps need not form a convex sequence, but among the convex sequences below them there is a pointwise largest one, their greatest convex minorant $S^\sharp$, which is the lower convex hull of the points $(s,\beta_s)$. Since the expected number of rejections, $S_1(p)+2\sum_{s\geq2}S_s(p)$, increases with every $S_s(p)$, we take the rule whose cumulative charges are $S^\sharp$; Proposition~\ref{prop:rebh-mean} shows that it satisfies all the other constraints in \eqref{eq:mean-menu} as well. Its tails are
	\begin{equation}\label{eq:mean-tails}
		q^\sharp_r=r\,S^\sharp_r-(r-1)\,S^\sharp_{r+1},\qquad r=1,\ldots,K.
	\end{equation}

	Geometrically, $q^\sharp_r$ is the value at $s=1$ of the line that extends the segment of the minorant over $[r,r+1]$; the same construction applied to the convex sequence $(\alpha/K)\sum_{j\geq s}e_{[j]}$ of cumulative budgets of $E^{\mathrm{eBH}}$ gives the bounds $q^*_r$ of \eqref{eq:rebh-tails} before capping. We call the resulting procedure \emph{recertified closed e-BH} (RC-eBH), since it is certified by the e-collection of closed e-BH. It replaces $q^*_r$ by $q^\sharp_r$ in Algorithm~\ref{alg:rebh}, that is, it rejects $R_{\kappa^\sharp}(e)$ with $\kappa^\sharp=\max(\{r:U\leq q^\sharp_r\}\cup\{0\})$. Since the points $(s,\beta_s)$ are ordered in $s$, the minorant is computed in $O(K)$ operations once the e-values are sorted.

	\begin{proposition}[Recertified closed e-BH]\label{prop:rebh-mean}
		RC-eBH has the following properties.
		\begin{enumerate}[label=(\roman*),leftmargin=*,itemsep=0pt]
			\item Its FDR is at most $\alpha$ in every state $T$ and under every law in $\Pc_T$.
			\item Its tails satisfy $q^\sharp_r\geq q^*_r$ for every $r$, so it contains the rejection set of RU-eBH, and hence those of U-eBH and e-BH, for every $e$ and $U$.
			\item Among all randomized rules that select from the certified menu of $E^{\overline{\mathrm{eBH}}}$, monotone or not, it has the largest expected number of rejections given $E=e$. In particular, it rejects at least as many hypotheses in expectation as closed e-BH.
		\end{enumerate}
	\end{proposition}
	The proof, which we defer to Appendix~\ref{app:rebh-mean}, shows that, for the charges of the minorant, the hardest constraints of each size are those for the sets of lowest-ranked hypotheses, so the constraints for these sets imply all others in \eqref{eq:mean-menu}. Unlike the menu of $E^{\mathrm{eBH}}$, the menu of the mean e-collection generally has no stochastically largest element, which is why part (iii) concerns the expected number of rejections. Whereas the FDR of RU-eBH is bounded by $\alpha|T|/K$, that of RC-eBH is bounded by $\alpha$.

	\begin{remark}[Conditional calibration]\label{rem:cc}
		Conditional calibration \citep{fithianlei2022,leeren2024} is another way of improving e-BH, and it follows the same pattern with a different source of budget. We only recast the e-BH version of \citet{leeren2024} in the present terms without improving or modifying it. Its baseline is the part of each e-value that e-BH actually uses. With $R_0$ the e-BH rejection set, the pruned e-value $h_j=K\ind\{j\in R_0\}/\{\alpha(|R_0|\vee1)\}\leq E_j$ leads e-BH to the same rejections. Suppose that under $H_j$ a statistic $S_j$ makes the conditional law of the data given $S_j$ known. The budget left unused by the pruned e-value, $\E[E_j-h_j\mid S_j]$, is then known, and \citet{leeren2024} spend it by raising $h_j$ to a larger e-value whose conditional expectation given $S_j$ does not exceed that of $E_j$. This is Proposition~\ref{prop:conditional} in e-value units, with conditioning statistic $S_j$ and budget $\E[E_j\mid S_j]$, and e-BH applied to the new e-values rejects at least $R_0$. Conditional calibration thus spends unused budget across the values of the data that share the same $S_j$, whereas RU-eBH spends it across the values of the external seed. Appendix~\ref{app:cc} gives the exact correspondence.
	\end{remark}

	\begin{remark}[Selective inference]\label{rem:cs}
		RU-eBH also applies to settings where the set of true nulls is allowed to be random. Every distribution in the menu \eqref{eq:rebh-budget} satisfies $\bar L(T,p;e)\leq(\alpha/K)\sum_{j\in T}e_j$ for every $T$, so with a seed independent of the data and of $T$, RU-eBH controls the FDR whenever $\E[E_j\ind\{j\in T\}]\leq1$ for every $j$. This is the setting of conformal selection under dependence \citep{jin2023wcs,bai2024optcs}. There, deterministic pruning is e-BH applied to such e-values, and the same calculation shows that pruning with a common uniform or with independent uniforms is U-eBH or J-eBH, respectively. RU-eBH applied to the same e-values therefore selects a superset of U-eBH for every realization of the uniform and, since J-eBH is also certified by $E^{\mathrm{eBH}}$, at least as many test points as J-eBH in expectation (Proposition~\ref{prop:rebh-general}). The argument does not extend to RC-eBH, because the mean e-collection \eqref{eq:mean-collection} normalizes by the random number of true nulls.
	\end{remark}

	\subsection{Simulations}\label{subsec:rebh-sim}
	We compare RU-eBH and RC-eBH with six procedures that, like them, control the FDR under arbitrary dependence. Four are deterministic: e-BH, its minimally adaptive version (MA-eBH) \citep{ignatiadis2024evalues}, closed e-BH \citep{xu2026closure} and anchored e-BH \citep{Ramdas2026Anchored}. Two are randomized: U-eBH \citep{xu2023randomization} and J-eBH (Section~\ref{subsec:rebh-slack}).

\begin{figure}[!p]
		\centering
		\includegraphics[width=\textwidth]{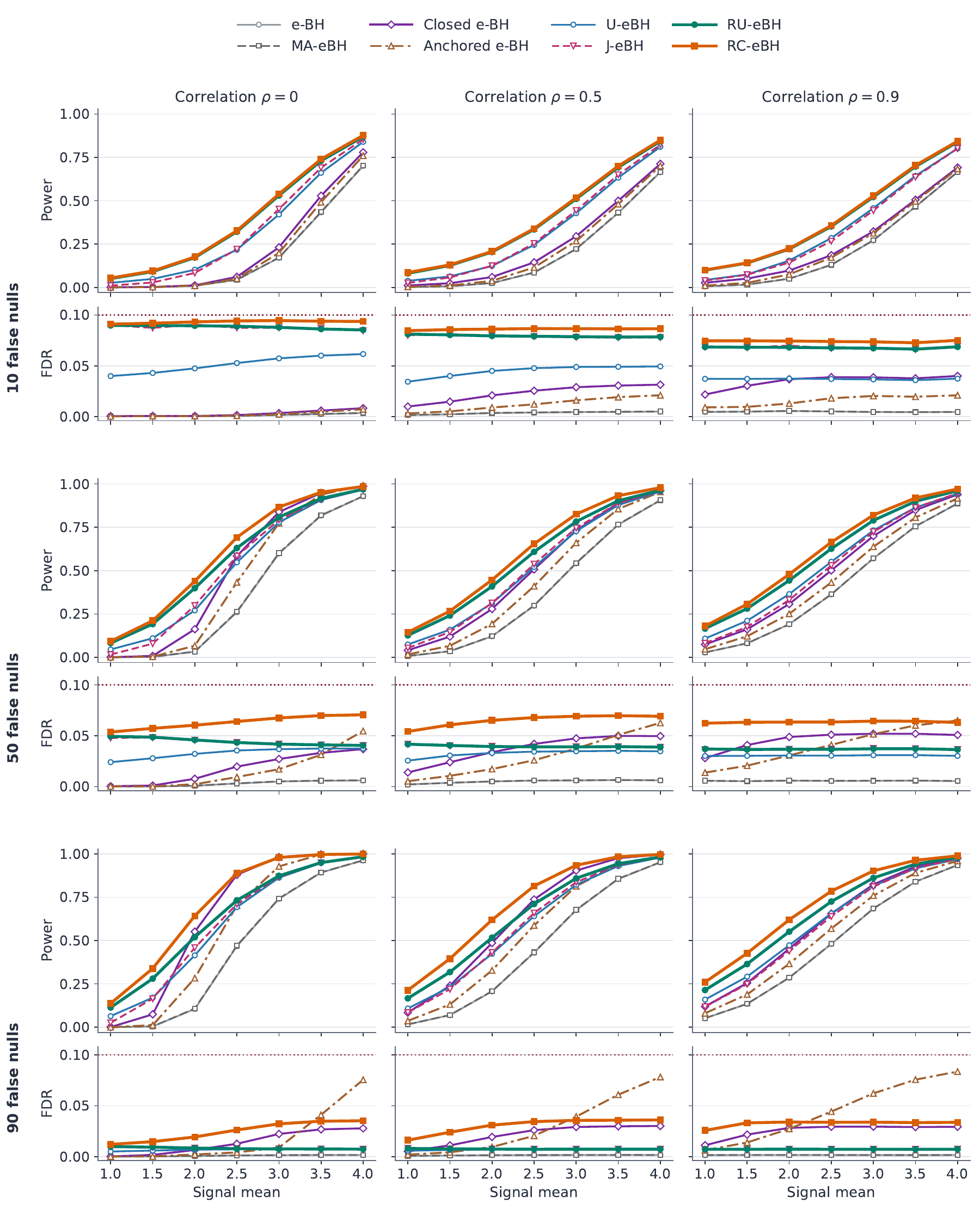}
		\caption{Power and FDR of RU-eBH, RC-eBH and six other procedures that are valid under arbitrary dependence, with $10$, $50$ and $90$ false nulls among $K=100$ hypotheses at $\alpha=0.1$ and three levels of correlation (columns). Each point averages $50{,}000$ data sets. The dotted lines mark $\alpha$.}
		\label{fig:rebh-sim}
	\end{figure}

	There are $K=100$ hypotheses, of which $10$, $50$ or $90$ are false. For each hypothesis we observe a Gaussian statistic $Z_j$ with unit variance, mean $0$ under the null and a common signal mean between $1$ and $4$ otherwise; the statistics are equicorrelated with correlation $\rho\in\{0,0.5,0.9\}$. All methods use the same e-values, the likelihood ratios $E_j=\exp(2Z_j-2)$ of $N(2,1)$ against $N(0,1)$, at level $\alpha=0.1$. We report power, the expected fraction of false nulls that are rejected, and the FDR, each averaged over $50{,}000$ data sets. Figure~\ref{fig:rebh-sim} shows all settings. The standard errors are very small, at most $0.0023$ for power and $0.0013$ for the FDR, so they are not shown.

	With $10$ false nulls, RU-eBH and RC-eBH are the most powerful methods in every panel, and they nearly coincide. At signal mean $2.5$ with independent statistics, for example, RU-eBH has power $0.32$, compared with $0.22$ for U-eBH and J-eBH and at most $0.06$ for the deterministic procedures. Its FDR stays close to the bound $\alpha|T|/K=0.09$ of Proposition~\ref{prop:rebh} with independent statistics, whereas the FDR of U-eBH is between about $40\%$ and $70\%$ of this bound, which reflects the budget that U-eBH leaves unused. J-eBH has nearly the same FDR as RU-eBH but much lower power, so the gain of RU-eBH also comes from how it allocates the budget rather than from how much of the budget it spends.

	With more false nulls, RU-eBH remains at least as powerful as U-eBH and J-eBH, but the deterministic procedures based on larger e-collections catch up. With $50$ false nulls and independent statistics, closed and anchored e-BH are more powerful than RU-eBH at strong signals, and with $90$ false nulls, closed e-BH is more powerful at moderate and strong signals unless the correlation is high. This is consistent with the theory: the optimality of RU-eBH is relative to $E^{\mathrm{eBH}}$, which normalizes by $K$, whereas closed e-BH is certified by the mean e-collection, which normalizes by $|T|$ and is much larger when few hypotheses are null. RC-eBH, which is certified by the same e-collection as closed e-BH, removes this gap. As part (ii) of Proposition~\ref{prop:rebh-mean} guarantees, it is more powerful than RU-eBH in every setting, and the gain grows with the number of false nulls; with $90$ false nulls, independent statistics and signal mean $2$, its power is $0.64$, compared with $0.52$ for RU-eBH and $0.55$ for closed e-BH.
	It is the most powerful of all eight methods,
	except at the strongest signals with independent statistics, where several methods reject nearly all false nulls and it trails the best of them by less than $0.005$. Its FDR stays below $\alpha$ but exceeds $\alpha|T|/K$ when many hypotheses are false, as the mean e-collection permits.

	\section{Discussion}\label{sec:utility}
	In both applications, recertification enlarges the menu of certified actions, and we selected its most liberal element -- the least conservative controller in CRC, and the boosting rule that rejects the most hypotheses in multiple testing. Nothing in the framework requires this choice. By Theorem~\ref{thm:closure}, every data-dependent selection from a certified menu is valid, so one may instead select the action that maximizes a utility, for instance a posterior expected utility under a prior. Such a selection is Bayes optimal among selections from the menu, and its validity does not depend on the prior. This differs from the constrained problems of \citet{wald1950} and \citet{hodgeslehmann1952}, which optimize a single rule subject to the risk constraints and, by Theorem~\ref{thm:closure}, amount to optimizing the e-collection and the selection jointly. Similarly, our admissibility results compare procedures on every calibration sample or conditionally on the e-values; they do not address classical risk admissibility, which allows a rule to trade larger losses at some observations for smaller losses at others.

	In summary, this paper formulates the e-closure principle for general decision problems and uses the resulting correspondence constructively. Recertification starts by measuring the budget that a valid procedure leaves unused, spends it through larger e-values, and inverts them into an improved procedure with the same guarantee. The source of this budget differs between our two applications. In conformal risk control, exchangeability makes the budget computable on every hypothetical augmented sample, and recertification yields pointwise improvements of CRC and an admissible limit. In multiple testing, the e-values themselves supply the budget once randomized procedures are treated as randomized actions, and recertifying U-eBH yields two closed-form procedures, RU-eBH and RC-eBH, each optimal among the procedures certified by its e-collection.

	The two applications also differ in cost. RU-eBH and RC-eBH require only sorting the e-values and, for RC-eBH, a convex minorant, whereas repeated recertification of CRC trades computation for a uniform improvement, with diminishing returns, and can be expensive when the controller depends on individual scores. Exploiting the structure of the loss and the data to compute stronger certificates economically remains an important practical question, and whether RC-eBH is admissible among all FDR procedures is open. More broadly, any procedure whose validity proof leaves an identifiable budget unused is a candidate for recertification, and we expect further applications of this kind, especially those with more complex states, actions or loss structures.

	\paragraph{Acknowledgements.}
	The main idea of the paper, recertifying a valid procedure by enlarging the e-values that certify it within its risk budget, was developed by the authors without AI. AI helped with coding and contributed to parts of the analysis, namely the admissibility of the limit of repeated recertification for convex losses in conformal risk control (Theorem~\ref{thm:crc-limit}) and the convex-minorant analysis of recertified closed e-BH (Section~\ref{subsec:rebh-mean}). The authors also used AI to revise the paper, and then edited it manually for clarity. The authors take responsibility for all content.

	\bibliographystyle{plainnat}
	\bibliography{reference}

\begin{thebibliography}{53}
\providecommand{\natexlab}[1]{#1}
\providecommand{\url}[1]{\texttt{#1}}
\expandafter\ifx\csname urlstyle\endcsname\relax
  \providecommand{\doi}[1]{doi: #1}\else
  \providecommand{\doi}{doi: \begingroup \urlstyle{rm}\Url}\fi

\bibitem[Andrews and Chen(2025)]{andrewschen2025}
I.~Andrews and J.~Chen.
\newblock Certified decisions.
\newblock \emph{arXiv preprint arXiv:2502.17830}, 2025.

\bibitem[Angelopoulos(2026)]{angelopoulos2026nonmonotonic}
A.~N. Angelopoulos.
\newblock Conformal risk control for non-monotonic losses.
\newblock \emph{arXiv preprint arXiv:2602.20151}, 2026.

\bibitem[Angelopoulos et~al.(2024)Angelopoulos, Bates, Fisch, Lei, and
  Schuster]{angelopoulos2024crc}
A.~N. Angelopoulos, S.~Bates, A.~Fisch, L.~Lei, and T.~Schuster.
\newblock Conformal risk control.
\newblock In \emph{International Conference on Learning Representations}, 2024.

\bibitem[Angelopoulos et~al.(2025)Angelopoulos, Bates, Cand{\`e}s, Jordan, and
  Lei]{angelopoulos2025ltt}
A.~N. Angelopoulos, S.~Bates, E.~J. Cand{\`e}s, M.~I. Jordan, and L.~Lei.
\newblock Learn then test: Calibrating predictive algorithms to achieve risk
  control.
\newblock \emph{The Annals of Applied Statistics}, 19\penalty0 (2):\penalty0
  1641--1662, 2025.

\bibitem[Bai and Jin(2024)]{bai2024optcs}
T.~Bai and Y.~Jin.
\newblock Optimized conformal selection: Powerful selective inference after
  conformity score optimization.
\newblock \emph{arXiv preprint arXiv:2411.17983}, 2024.

\bibitem[Bai and Jin(2026)]{bai2026score}
T.~Bai and Y.~Jin.
\newblock Conformal selective prediction with general risk control.
\newblock \emph{arXiv preprint arXiv:2603.24704}, 2026.

\bibitem[Bates et~al.(2021)Bates, Angelopoulos, Lei, Malik, and
  Jordan]{bates2021rcps}
S.~Bates, A.~N. Angelopoulos, L.~Lei, J.~Malik, and M.~I. Jordan.
\newblock Distribution-free, risk-controlling prediction sets.
\newblock \emph{Journal of the ACM}, 68\penalty0 (6):\penalty0 Article 43,
  2021.

\bibitem[Benjamini and Hochberg(1995)]{benjaminihochberg1995}
Y.~Benjamini and Y.~Hochberg.
\newblock Controlling the false discovery rate: A practical and powerful
  approach to multiple testing.
\newblock \emph{Journal of the Royal Statistical Society: Series B},
  57\penalty0 (1):\penalty0 289--300, 1995.

\bibitem[Berger(1985)]{berger1985}
J.~O. Berger.
\newblock \emph{Statistical Decision Theory and {B}ayesian Analysis}.
\newblock Springer, second edition, 1985.

\bibitem[Blanchard and Roquain(2008)]{blanchard2008two}
G.~Blanchard and E.~Roquain.
\newblock Two simple sufficient conditions for {FDR} control.
\newblock \emph{Electronic Journal of Statistics}, 2:\penalty0 963--992, 2008.

\bibitem[Chugg et~al.(2026{\natexlab{a}})Chugg, Jordan, and
  Ramdas]{chugg2026strategic}
B.~Chugg, M.~I. Jordan, and A.~Ramdas.
\newblock Strategic multi-agent resource allocation via statistical contracts.
\newblock \emph{arXiv preprint arXiv:2610.18336}, 2026{\natexlab{a}}.

\bibitem[Chugg et~al.(2026{\natexlab{b}})Chugg, Lardy, Ramdas, and
  Gr{\"u}nwald]{chugg2026admissibility}
B.~Chugg, T.~Lardy, A.~Ramdas, and P.~D. Gr{\"u}nwald.
\newblock On admissibility in post-hoc hypothesis testing.
\newblock \emph{International Journal of Approximate Reasoning}, 191:\penalty0
  109634, 2026{\natexlab{b}}.

\bibitem[Cohen et~al.(2024)Cohen, Park, Simeone, and Shamai]{cohen2024cross}
K.~M. Cohen, S.~Park, O.~Simeone, and S.~Shamai.
\newblock Cross-validation conformal risk control.
\newblock In \emph{IEEE International Symposium on Information Theory}, 2024.

\bibitem[Farinhas et~al.(2024)Farinhas, Zerva, Ulmer, and
  Martins]{farinhas2024nonexchangeable}
A.~Farinhas, C.~Zerva, D.~Ulmer, and A.~F.~T. Martins.
\newblock Non-exchangeable conformal risk control.
\newblock In \emph{International Conference on Learning Representations}, 2024.

\bibitem[Fithian and Lei(2022)]{fithianlei2022}
W.~Fithian and L.~Lei.
\newblock Conditional calibration for false discovery rate control under
  dependence.
\newblock \emph{The Annals of Statistics}, 50\penalty0 (6):\penalty0
  3091--3118, 2022.

\bibitem[Gauthier et~al.(2025)Gauthier, Bach, and Jordan]{gauthier2025evalues}
E.~Gauthier, F.~Bach, and M.~I. Jordan.
\newblock E-values expand the scope of conformal prediction.
\newblock \emph{arXiv preprint arXiv:2503.13050}, 2025.

\bibitem[Goeman(2026)]{Goeman2026Closed}
J.~J. Goeman.
\newblock A uniform improvement of the {B}enjamini--{H}ochberg procedure via
  e-closure.
\newblock \emph{arXiv preprint arXiv:2606.01854}, 2026.

\bibitem[Goeman and Solari(2011)]{goemansolari2011}
J.~J. Goeman and A.~Solari.
\newblock Multiple testing for exploratory research.
\newblock \emph{Statistical Science}, 26\penalty0 (4):\penalty0 584--597, 2011.

\bibitem[Goeman et~al.(2021)Goeman, Hemerik, and Solari]{goeman2021only}
J.~J. Goeman, J.~Hemerik, and A.~Solari.
\newblock Only closed testing procedures are admissible for controlling false
  discovery proportions.
\newblock \emph{The Annals of Statistics}, 49\penalty0 (2):\penalty0
  1218--1238, 2021.

\bibitem[Gonz{\'a}lez-Gonz{\'a}lez(2026)]{gonzalez2026embedded}
P.~Gonz{\'a}lez-Gonz{\'a}lez.
\newblock Embedded representations of popular image datasets ({CIFAR10},
  {CIFAR100}, {CIFAR100coarse}, {MNIST}, {FashionMNIST}, {SVHN}).
\newblock Zenodo dataset, version 1,
  \url{https://doi.org/10.5281/zenodo.21131944}, 2026.

\bibitem[Gr{\"u}nwald(2023)]{grunwald2023eposterior}
P.~D. Gr{\"u}nwald.
\newblock The e-posterior.
\newblock \emph{Philosophical Transactions of the Royal Society A},
  381\penalty0 (2247):\penalty0 20220146, 2023.

\bibitem[Gr{\"u}nwald(2024)]{grunwald2024beyond}
P.~D. Gr{\"u}nwald.
\newblock Beyond {N}eyman--{P}earson: E-values enable hypothesis testing with a
  data-driven alpha.
\newblock \emph{Proceedings of the National Academy of Sciences}, 121\penalty0
  (39):\penalty0 e2302098121, 2024.

\bibitem[Hodges and Lehmann(1952)]{hodgeslehmann1952}
J.~L. Hodges, Jr. and E.~L. Lehmann.
\newblock The use of previous experience in reaching statistical decisions.
\newblock \emph{The Annals of Mathematical Statistics}, 23\penalty0
  (3):\penalty0 396--407, 1952.

\bibitem[Ignatiadis et~al.(2024)Ignatiadis, Wang, and
  Ramdas]{ignatiadis2024evalues}
N.~Ignatiadis, R.~Wang, and A.~Ramdas.
\newblock E-values as unnormalized weights in multiple testing.
\newblock \emph{Biometrika}, 111\penalty0 (2):\penalty0 417--439, 2024.

\bibitem[Jin and Cand{\`e}s(2023)]{jin2023wcs}
Y.~Jin and E.~J. Cand{\`e}s.
\newblock Model-free selective inference under covariate shift via weighted
  conformal p-values.
\newblock \emph{arXiv preprint arXiv:2307.09291}, 2023.

\bibitem[Katsevich and Ramdas(2020)]{katsevich2020simultaneous}
E.~Katsevich and A.~Ramdas.
\newblock Simultaneous high-probability bounds on the false discovery
  proportion in structured, regression and online settings.
\newblock \emph{The Annals of Statistics}, 48\penalty0 (6):\penalty0
  3465--3487, 2020.

\bibitem[Kiyani et~al.(2025)Kiyani, Pappas, Roth, and
  Hassani]{kiyani2025decision}
S.~Kiyani, G.~J. Pappas, A.~Roth, and H.~Hassani.
\newblock Decision theoretic foundations for conformal prediction: Optimal
  uncertainty quantification for risk-averse agents.
\newblock In \emph{Proceedings of the 42nd International Conference on Machine
  Learning}, volume 267 of \emph{Proceedings of Machine Learning Research},
  pages 30943--30965, 2025.

\bibitem[Koning(2025)]{koning2023posthoc}
N.~W. Koning.
\newblock Post-hoc $\alpha$ hypothesis testing and the post-hoc $p$-value.
\newblock \emph{arXiv preprint arXiv:2312.08040}, 2025.

\bibitem[Koning(2026{\natexlab{a}})]{koning2026axioms}
N.~W. Koning.
\newblock Axioms for testing with data-dependent levels, e-values and p-values.
\newblock \emph{arXiv preprint arXiv:2605.28429}, 2026{\natexlab{a}}.

\bibitem[Koning(2026{\natexlab{b}})]{koning2026emeasure}
N.~W. Koning.
\newblock The {E}-measure.
\newblock \emph{arXiv preprint arXiv:2604.20788}, 2026{\natexlab{b}}.

\bibitem[Koning(2026{\natexlab{c}})]{koning2026invariance}
N.~W. Koning.
\newblock Measuring evidence against exchangeability and group invariance with
  e-values.
\newblock \emph{The Annals of Statistics}, 54\penalty0 (4):\penalty0
  1843--1869, 2026{\natexlab{c}}.

\bibitem[Koning and van Meer(2026)]{koningvanmeer2026}
N.~W. Koning and S.~van Meer.
\newblock Fuzzy prediction sets: Conformal prediction with e-values.
\newblock \emph{arXiv preprint arXiv:2509.13130}, 2026.

\bibitem[Lee and Ren(2024)]{leeren2024}
J.~Lee and Z.~Ren.
\newblock Boosting e-{BH} via conditional calibration.
\newblock \emph{arXiv preprint arXiv:2404.17562}, 2024.

\bibitem[Lehmann and Romano(2022)]{lehmannromano2022}
E.~L. Lehmann and J.~P. Romano.
\newblock \emph{Testing Statistical Hypotheses}.
\newblock Springer, fourth edition, 2022.

\bibitem[Lekeufack et~al.(2024)Lekeufack, Angelopoulos, Bajcsy, Jordan, and
  Malik]{lekeufack2024conformal}
J.~Lekeufack, A.~N. Angelopoulos, A.~Bajcsy, M.~I. Jordan, and J.~Malik.
\newblock Conformal decision theory: Safe autonomous decisions from imperfect
  predictions.
\newblock In \emph{IEEE International Conference on Robotics and Automation},
  2024.

\bibitem[Marcus et~al.(1976)Marcus, Peritz, and Gabriel]{marcus1976}
R.~Marcus, E.~Peritz, and K.~R. Gabriel.
\newblock On closed testing procedures with special reference to ordered
  analysis of variance.
\newblock \emph{Biometrika}, 63\penalty0 (3):\penalty0 655--660, 1976.

\bibitem[Ramdas(2026)]{Ramdas2026Anchored}
A.~Ramdas.
\newblock Anchored multiple testing: A transparent use of e-closure to improve
  {FDR} procedures.
\newblock \emph{arXiv preprint arXiv:2610.11279}, 2026.

\bibitem[Ramdas and Wang(2025)]{ramdaswang2025}
A.~Ramdas and R.~Wang.
\newblock Hypothesis testing with e-values.
\newblock \emph{Foundations and Trends in Statistics}, 1\penalty0
  (1--2):\penalty0 1--390, 2025.

\bibitem[Solari and Goeman(2017)]{solari2017minimally}
A.~Solari and J.~J. Goeman.
\newblock Minimally adaptive {BH}: A tiny but uniform improvement of the
  procedure of {B}enjamini and {H}ochberg.
\newblock \emph{Biometrical Journal}, 59\penalty0 (4):\penalty0 776--780, 2017.

\bibitem[Sonnemann(2008)]{sonnemann2008}
E.~Sonnemann.
\newblock General solutions to multiple testing problems.
\newblock \emph{Biometrical Journal}, 50\penalty0 (5):\penalty0 641--656, 2008.

\bibitem[Srivastava(1998)]{srivastava1998}
S.~M. Srivastava.
\newblock \emph{A Course on {B}orel Sets}, volume 180 of \emph{Graduate Texts
  in Mathematics}.
\newblock Springer, 1998.

\bibitem[Sun and Wang(2026)]{sunwang2026}
L.~Sun and R.~Wang.
\newblock Admissibility and complete classes for false discovery rate control
  with e-values.
\newblock \emph{arXiv preprint arXiv:2607.14380}, 2026.

\bibitem[Tibshirani et~al.(2026)Tibshirani, Barber, and
  Ramdas]{tibshirani2026lens}
R.~J. Tibshirani, R.~F. Barber, and A.~Ramdas.
\newblock Conformal prediction through the lens of hypothesis testing:
  Universality, impossibility, and optimality.
\newblock \emph{arXiv preprint arXiv:2608.27310}, 2026.

\bibitem[Vovk(2025)]{vovk2025eprediction}
V.~Vovk.
\newblock Conformal e-prediction.
\newblock \emph{Pattern Recognition}, 166:\penalty0 111674, 2025.

\bibitem[Vovk and Wang(2021)]{vovkwang2021}
V.~Vovk and R.~Wang.
\newblock E-values: Calibration, combination and applications.
\newblock \emph{The Annals of Statistics}, 49\penalty0 (3):\penalty0
  1736--1754, 2021.

\bibitem[Vovk et~al.(2022)Vovk, Gammerman, and Shafer]{vovk2022algorithmic}
V.~Vovk, A.~Gammerman, and G.~Shafer.
\newblock \emph{Algorithmic Learning in a Random World}.
\newblock Springer, second edition, 2022.

\bibitem[Wald(1950)]{wald1950}
A.~Wald.
\newblock \emph{Statistical Decision Functions}.
\newblock John Wiley \& Sons, 1950.

\bibitem[Wang and Ramdas(2022)]{wangramdas2022}
R.~Wang and A.~Ramdas.
\newblock False discovery rate control with e-values.
\newblock \emph{Journal of the Royal Statistical Society: Series B},
  84\penalty0 (3):\penalty0 822--852, 2022.

\bibitem[Xu and Ramdas(2023)]{xu2023randomization}
Z.~Xu and A.~Ramdas.
\newblock More powerful multiple testing under dependence via randomization.
\newblock \emph{arXiv preprint arXiv:2305.11126}, 2023.

\bibitem[Xu et~al.(2024)Xu, Wang, and Ramdas]{xu2024postselection}
Z.~Xu, R.~Wang, and A.~Ramdas.
\newblock Post-selection inference for e-value based confidence intervals.
\newblock \emph{Electronic Journal of Statistics}, 18\penalty0 (1):\penalty0
  2292--2338, 2024.

\bibitem[Xu et~al.(2026{\natexlab{a}})Xu, Fischer, and Ramdas]{xu2026online}
Z.~Xu, L.~Fischer, and A.~Ramdas.
\newblock Improving online {FDR} procedures via online analogs of e-closure and
  compound e-values.
\newblock In \emph{Proceedings of the Conference on Uncertainty in Artificial
  Intelligence}, volume 337 of \emph{Proceedings of Machine Learning Research},
  pages 7552--7561, 2026{\natexlab{a}}.

\bibitem[Xu et~al.(2026{\natexlab{b}})Xu, Solari, Fischer, de~Heide, Ramdas,
  and Goeman]{xu2026closure}
Z.~Xu, A.~Solari, L.~Fischer, R.~de~Heide, A.~Ramdas, and J.~Goeman.
\newblock Bringing closure to false discovery rate control: A general principle
  for multiple testing.
\newblock \emph{arXiv preprint arXiv:2509.02517}, 2026{\natexlab{b}}.

\bibitem[Yeh et~al.(2025)Yeh, Christianson, Wierman, and Yue]{yeh2025training}
C.~Yeh, N.~Christianson, A.~Wierman, and Y.~Yue.
\newblock Conformal risk training: End-to-end optimization of conformal risk
  control.
\newblock In \emph{Advances in Neural Information Processing Systems}, 2025.

\end{thebibliography}
	\clearpage

	\appendix
	\section{Measurability of the certified menu}\label{app:measurable}
	The proof of Theorem~\ref{thm:closure} fixes one state--loss pair before taking expectations, so it requires only that the rule and the losses be measurable. The certified menu is different: it intersects possibly uncountably many constraints, which in general need not produce a Borel graph, and selecting from it requires measurable selections. The following conditions are sufficient, though not necessary.

	\begin{proposition}\label{prop:regularity}
		Suppose that $\Xc$ is standard Borel, $\Sc$ and $\mathcal T$ are separable metric spaces, and $\Ac$ is a compact metric space. Suppose that the losses and the e-collection are finite, that $\alpha_t(s)>0$, and that both $L_t(s,a;x)$ and
		\[
		f(x,s,t,a)=L_t(s,a;x)-\alpha_t(s)E_{s,t}(x)
		\]
		are Borel in $x$ for each fixed $(s,t,a)$ and continuous in $(s,t,a)$ for each fixed $x$. If the certified menu is nonempty for every $x$, then it has a Borel graph and nonempty compact values, and it admits a Borel selection. A utility $v(x,a)$ that is Borel in $x$ and continuous in $a$ has a Borel maximizing selection from the menu. Every loss envelope of the menu is Borel in $x$ and continuous in $(s,t)$.
	\end{proposition}
	\begin{proof}
		Choose a countable dense set $\{(s_j,t_j):j\geq1\}$ in $\Sc\times\mathcal T$. By continuity,
		\[
		\Cc_\alpha(E;x)=\bigcap_{j\geq1}\{a:f(x,s_j,t_j,a)\leq0\}.
		\]
		Each constraint is jointly Borel in $(x,a)$, so the graph is Borel, and its sections are closed, hence compact. The Arsenin--Kunugui theorem \citep[Theorem 5.12.1]{srivastava1998} then gives a Borel selection. For a utility $v$, the sets $\{(x,a):a\in\Cc_\alpha(E;x),\ v(x,a)>b\}$ have $\sigma$-compact sections, so their projections are Borel by the same theorem, and the maximum value is Borel in $x$. The maximizing graph is therefore Borel with nonempty compact sections, and a further application gives a maximizing selection. The envelopes are Borel by the same argument applied to each fixed-pair loss, and continuous in $(s,t)$ as maxima of jointly continuous functions over the compact set $\Cc_\alpha(E;x)$.
	\end{proof}

	\section{Details for conformal risk control}\label{app:crc}
	\subsection{CRC through e-closure} \label{app:crc-baseline}
	We work in the formulation of Section~\ref{subsec:crc-refine}, in which a state is an augmented multiset $S=(z_1,\ldots,z_{n+1})$ and $\Pc_S$ deletes a uniformly random point of $S$.
	For a fixed action $\lambda$, the losses $\ell(z_i,\lambda)$ act as scores of the points of $S$, and
	\[
	E^\lambda_S(D)=\frac{(n+1)\,\ell(S\setminus D,\lambda)}{\sum_{i=1}^{n+1}\ell(z_i,\lambda)}\qquad(D\subseteq S),
	\]
	with $0/0=0$, is the conformal e-value of the held-out point \citep{vovk2025eprediction,koning2026invariance}. Its expectation under $\Pc_S$ is at most one, so $E^\lambda$ is an e-collection for each fixed $\lambda$.

	Theorem~\ref{thm:closure} certifies all actions with a single e-collection. Here we instead allow each action $\lambda$ to be certified by its own e-values $E^\lambda$, and consider, at an observed $D$, the actions that satisfy
	\begin{equation}\label{eq:own-evalues}
		\ell(z,\lambda)\leq\alpha E^\lambda_{D\cup\{z\}}(D)\qquad\text{for every }z\in\Zc.
	\end{equation}
	The constraint is trivial when $\ell(z,\lambda)=0$, and otherwise it reads $\sum_{i=1}^n\ell(Z_i,\lambda)+\ell(z,\lambda)\leq(n+1)\alpha$. Hence \eqref{eq:own-evalues} holds if and only if
	\[
	\sum_{i=1}^n\ell(Z_i,\lambda)+U^*(\lambda)\leq(n+1)\alpha,\qquad U^*(\lambda)=\sup_{z\in\Zc}\ell(z,\lambda),
	\]
	which is the envelope CRC test \eqref{eq:env} with the sharp envelope $U^*$, non-increasing and right-continuous as a supremum of such functions. A larger envelope $U$, or the constant $B$ of standard CRC, replaces the unobserved test loss $\ell(z,\lambda)$ in the denominator of $E^\lambda$ by an upper bound and gives a more stringent test. Envelope CRC and CRC therefore also select actions that satisfy \eqref{eq:own-evalues}.

	In general, letting the e-values depend on the action is not valid: choosing the action after seeing the data then amounts to choosing among e-values, which inflates them. The conformal case avoids this due to a cancellation. Under $\Pc_S$ the augmented sample $S$ is fixed, and the data reveal only which of its points is held out. For a held-out point with positive loss, the constraint in \eqref{eq:own-evalues} reads $\sum_{j=1}^{n+1}\ell(z_j,\lambda)\leq(n+1)\alpha$, whichever point $z$ is held out. Whether an action passes is thus decided by the state alone, and the data cannot be used to choose among the e-values. Formally, let
	\[
	\lambda^*(S)=\min\left\{\lambda\in\Lambda:\sum_{j=1}^{n+1}\ell(z_j,\lambda)\leq (n+1)\alpha\right\}
	\]
	be the smallest action that passes on $S$.
	Since $\lambda^*(S)$ is permutation invariant in $S$, the values $E^*_S(D)=E^{\lambda^*(S)}_S(D)$ form a single e-collection. If $\lambda$ satisfies \eqref{eq:own-evalues} at $D$ and $\ell(z,\lambda)>0$, then $\lambda\geq\lambda^*(S)$ for $S=D\cup\{z\}$, and hence
	\[
	\ell(z,\lambda)\leq\ell(z,\lambda^*(S))\leq\alpha E^*_S(D),
	\]
	where the second inequality uses $\sum_j\ell(z_j,\lambda^*(S))\leq(n+1)\alpha$. Every action that satisfies \eqref{eq:own-evalues} therefore lies in the certified menu of $E^*$, and Theorem~\ref{thm:closure} gives the validity of CRC and envelope CRC.

	The e-collection $E^*$ is obtained by first choosing an action through risk control on the augmented sample and then evaluating conformal e-values at that action. The e-values of \citet{bai2026score} for selective prediction are built in the same way. There, a selection threshold is chosen by controlling the risk on the augmented sample, and the conformal e-value is evaluated at that threshold, with the unknown test loss handled by an infimum over its possible values, much as \eqref{eq:own-evalues} is imposed for every possible test point.

	\paragraph{Exchangeability and i.i.d. sampling.} \label{app:iid}
	Finally, we explain why Lemma~\ref{lem:orbit} requires validity under every finite exchangeable law.
	Validity under i.i.d. sampling extends to mixtures of i.i.d. laws, but finite exchangeable laws need not be such mixtures, and the two requirements already differ for $n=1$. Let $Z_1,Z_2\in\{0,1\}$, take the miscoverage loss with $\alpha=1/2$, and consider the controller that predicts the set $\{Z_1\}$ for $Z_2$. If $Z_1$ and $Z_2$ are i.i.d. Bernoulli$(p)$, it miscovers with probability $2p(1-p)\leq1/2$, so it is valid under i.i.d. sampling. Under the exchangeable law that is uniform on $(0,1)$ and $(1,0)$, which is not a mixture of i.i.d. laws, it miscovers with probability one. In terms of Lemma~\ref{lem:orbit}, both held-out losses on the augmented sample $\{0,1\}$ equal one, so their total exceeds the budget $(n+1)\alpha=1$ and the unused budget $\Delta_A(\{0,1\})=-1$ is negative. Recertification spends $\Delta_A$ and therefore needs validity under exchangeability. CRC and envelope CRC have it, and so satisfy \eqref{eq:orbit-characterization}.

	\subsection{Other allocations}\label{app:allocations}

	Equal sharing is one way, but not the only way, to distribute the unused budget among the held-out positions. For each augmented multiset $S$ and risk $t$, let $w_{i,t}(S)\geq0$ be measurable weights with $\sum_iw_{i,t}(S)=1$ that are permutation equivariant: reordering the points of $S$ reorders the weights, and identical points receive identical weights. These weights give the budgets
	\begin{equation}\label{eq:alloc-general}
		b_{i,t}^A(S)=\ell_t(z_i,A(S_{-i}))+\Delta_{A,t}(S)\,w_{i,t}(S),
	\end{equation}
	which sum to $(n+1)\alpha_t$ on every augmented sample. Write $b^A_t(S;z)$ for the common budget of the copies of $z$ in $S$. By Lemma~\ref{lem:orbit}, every measurable symmetric controller $C$ with
	\begin{equation}\label{eq:alloc-constraint}
		\ell_t(z,C(D))\leq b^A_t(D\cup\{z\};z)\qquad\text{for all }D,\ z,\ t
	\end{equation}
	is jointly valid, since its held-out losses on each $S$ sum to at most $(n+1)\alpha_t$. Adding the constraint $C\leq A$ makes the improvement pointwise. Equal weights recover \eqref{eq:multi-recycle}, and weights proportional to the held-out losses of $A$ are another choice. Restricting to allocations loses no generality, since every pointwise improvement satisfies \eqref{eq:alloc-constraint} for some allocation.

	\begin{proposition}[Every pointwise improvement has an allocation]\label{prop:allocation-complete}
		Suppose that every $\ell_t(z,\cdot)$ is non-increasing. If $A$ and $C$ are jointly valid symmetric controllers with $C\leq A$, then $C$ satisfies \eqref{eq:alloc-constraint} for some permutation-equivariant weights.
	\end{proposition}
	\begin{proof}
		Fix $S$ and $t$, and write $r^A_{i,t}$ and $r^C_{i,t}$ for the held-out losses of the two controllers. Since $C\leq A$ and the losses are non-increasing, the extra losses $d_{i,t}=r_{i,t}^C-r_{i,t}^A$ are nonnegative, and they sum to $\Delta_{A,t}-\Delta_{C,t}$. If $\Delta_{A,t}>0$, set
		\[
		w_{i,t}(S)=\frac{d_{i,t}+\Delta_{C,t}/(n+1)}{\Delta_{A,t}}.
		\]
		These weights are nonnegative and sum to one. They give each position its extra loss together with an equal share of the unused budget of $C$, so the budgets in \eqref{eq:alloc-general} equal $r_{i,t}^C+\Delta_{C,t}/(n+1)\geq r_{i,t}^C$. If $\Delta_{A,t}=0$, then $\sum_id_{i,t}=-\Delta_{C,t}\leq0$ forces every $d_{i,t}$ to vanish, and any equivariant weights suffice. The weights are equivariant and measurable because both controllers are symmetric and measurable.
	\end{proof}

	By Proposition~\ref{prop:allocation-complete} and Theorem~\ref{thm:crc-limit}, repeated equal sharing can also be interpreted as an implicit search over allocations. Its admissible limit $A^{(\infty)}$ satisfies \eqref{eq:alloc-constraint} for some allocation of the unused budget of $A^{(0)}$, and no allocation yields a valid controller that improves on it.

	\subsection{Proof of Theorem~\ref{thm:crc-limit} and Proposition~\ref{prop:multi-risk}}\label{app:limit}
	We prove the second part of Proposition~\ref{prop:multi-risk}, of which Theorem~\ref{thm:crc-limit} is the case of a single risk. Throughout, $\Lambda=[\lambda_-,\lambda_+]$, and every $\ell_t(z,\cdot)$ is non-increasing, right-continuous and convex. By Lemma~\ref{lem:orbit}, a controller $A$ is jointly valid if and only if $T_{A,t}(S)\leq(n+1)\alpha_t$ for every $S$ and $t$.

	\begin{lemma}[One-step comparison]\label{lem:one-step}
		If $A$ and $C$ are jointly valid and $C\leq A$, then $\rec A\leq\rho A+C/(n+1)$.
	\end{lemma}
	\begin{proof}
		Fix $D$, $z$ and $t$, and let $S=D\cup\{z\}$. Since $C\leq A$ and the losses are non-increasing, each held-out loss of $C$ on $S$ is at least that of $A$, and by joint validity of $C$ these differences sum to $T_{C,t}(S)-T_{A,t}(S)\leq\Delta_{A,t}(S)$. Each difference is therefore at most $\Delta_{A,t}(S)$, and for the held-out point $z$,
		\[
		\ell_t(z,C(D))-\ell_t(z,A(D))\leq\Delta_{A,t}(S).
		\]
		The action $a=\rho A(D)+C(D)/(n+1)$ lies between $C(D)$ and $A(D)$, and convexity gives
		\[
		\ell_t(z,a)-\ell_t(z,A(D))
		\leq\frac{\ell_t(z,C(D))-\ell_t(z,A(D))}{n+1}
		\leq\frac{\Delta_{A,t}(S)}{n+1}.
		\]
		Since $z$ and $t$ are arbitrary, $a$ is feasible in \eqref{eq:multi-recycle}, and hence $(\rec A)(D)\leq a$.
	\end{proof}

	\begin{proof}[Proof of Proposition~\ref{prop:multi-risk}]
		By the first part of the proposition, the iterates are jointly valid and $A^{(k+1)}\leq A^{(k)}$. Hence $A^{(k)}(D)$ decreases to a limit $A^{(\infty)}(D)\in\Lambda$ for every $D$, and $A^{(\infty)}$ is measurable and symmetric as a pointwise limit. For fixed $S$ and $t$, the losses are non-increasing and right-continuous, so $\ell_t(z_i,A^{(k)}(S_{-i}))\uparrow\ell_t(z_i,A^{(\infty)}(S_{-i}))$, and $T_{A^{(\infty)},t}(S)=\lim_kT_{A^{(k)},t}(S)\leq(n+1)\alpha_t$. Thus $A^{(\infty)}$ is jointly valid.

		Let $C$ be jointly valid with $C\leq A^{(\infty)}$. Then $C\leq A^{(k)}$ for every $k$, and Lemma~\ref{lem:one-step} gives $A^{(k+1)}\leq\rho A^{(k)}+C/(n+1)$. Letting $k\to\infty$ yields $A^{(\infty)}\leq\rho A^{(\infty)}+C/(n+1)$, that is, $A^{(\infty)}\leq C$. Hence $C=A^{(\infty)}$, and no jointly valid controller strictly dominates $A^{(\infty)}$. Taking $C=A^{(\infty)}$ in Lemma~\ref{lem:one-step} gives $0\leq A^{(k+1)}-A^{(\infty)}\leq\rho\,(A^{(k)}-A^{(\infty)})$, and induction gives \eqref{eq:crc-rate}.

		Finally, if $A$ is admissible, then $\rec A$ is jointly valid with $\rec A\leq A$, so $\rec A=A$ and $A$ is a fixed point. Conversely, if $\rec A=A$ and $C\leq A$ is jointly valid, Lemma~\ref{lem:one-step} gives $A\leq\rho A+C/(n+1)$, so $A\leq C$ and $C=A$. Hence $A$ is admissible.
	\end{proof}

	\subsection{Proof of Proposition~\ref{prop:coordinate}}\label{app:coordinate}
	For a calibration multiset $D_0$ and an action $a\in\Lambda$, let $A^{D_0,a}$ denote the controller that agrees with $A$ except at $D_0$, where it takes the action $a$. It is symmetric, and measurable since it differs from $A$ at a single multiset.

	\begin{proof}[Proof of Proposition~\ref{prop:coordinate}]
		Consider changing the controller $A$ to $A^{D_0, a}$. The held-out position $i$ of an augmented multiset $S$ is affected by the change if and only if $S_{-i}=D_0$, that is, if and only if $S=D_0\cup\{z\}$ and $z_i=z$ for some $z$. Such an $S$ contains $\nu_{D_0}(z)+1$ copies of $z$, so
		\[
		T_{A^{D_0,a},t}(D_0\cup\{z\})=T_{A,t}(D_0\cup\{z\})+(\nu_{D_0}(z)+1)\{\ell_t(z,a)-\ell_t(z,A(D_0))\},
		\]
		and every other total is unchanged. By Lemma~\ref{lem:orbit}, $A^{D_0,a}$ is jointly valid if and only if \eqref{eq:coordinate-crc} holds at $D_0$ and $a$.
		If \eqref{eq:coordinate-crc} holds at some $D_0$ and $a<A(D_0)$, then $A^{D_0,a}$ is jointly valid and strictly dominates $A$. Conversely, let $C\leq A$ be a jointly valid symmetric controller with $C(D_0)<A(D_0)$ for some $D_0$, and put $a=C(D_0)$. Since $A^{D_0,a}\geq C$ pointwise and the losses are non-increasing, each held-out loss of $A^{D_0,a}$ is at most that of $C$. Hence $A^{D_0,a}$ is jointly valid, and \eqref{eq:coordinate-crc} holds at $D_0$ and $a<A(D_0)$.

		Since $\Zc$ is finite in this case, there are finitely many calibration multisets. Their order is fixed in advance, so that the sweep defines a single controller rather than a data-dependent procedure. At each visit, the actions satisfying \eqref{eq:coordinate-crc} form a closed upper interval of $\Lambda$ that contains the current entry, by the monotonicity and right continuity of the losses, so the smallest such action $a^*$ exists, and the update preserves joint validity. Two observations show that a visited entry cannot be lowered afterwards. First, right after the update, an action $a<a^*$ satisfies \eqref{eq:coordinate-crc} at $D_0$ if and only if it did before the update, since the update lowers both sides of \eqref{eq:coordinate-crc} by the same amount $(\nu_{D_0}(z)+1)\{\ell_t(z,a^*)-\ell_t(z,A(D_0))\}$. By the choice of $a^*$, no such action exists. Second, a later update at another multiset can only increase held-out losses, so it shrinks every unused budget, and hence the right side of \eqref{eq:coordinate-crc} at $D_0$, while the entry at $D_0$, and hence the left side, stays fixed. After one pass, the condition therefore fails at every multiset for every smaller action. The resulting controller is jointly valid and no larger than $A$, and it is admissible by the criterion.
	\end{proof}

	\subsection{The two-type example}\label{app:toy}
	In this subsection, we carefully solve the two-type example in Section~\ref{subsec:iteration}.

	We first write out the validity constraints, which describe the valid controllers as the points of $[0,1]^3$ that satisfy four linear inequalities.
	With the losses \eqref{eq:toy-loss}, $\ell(z,\lambda)=g(z)(1-\lambda)$ with $g(H)=1$ and $g(L)=1/2$, and the budget is $(n+1)\alpha=3/2$. A symmetric controller $A$ is determined by its actions at $HH$, $HL$ and $LL$. We write
	\[
	(h,u,v)=\bigl(1-A(HH),\,1-A(HL),\,1-A(LL)\bigr)\in[0,1]^3,
	\]
	so that larger coordinates correspond to less conservative actions. In $HHL$, for instance, deleting either copy of $H$ leaves $HL$ and deleting $L$ leaves $HH$, so the held-out losses are $u$, $u$ and $h/2$. Treating the four augmented multisets in this way, Lemma~\ref{lem:orbit} shows that $A$ is valid if and only if
	\begin{equation}\label{eq:toy-valid}
		3h\leq\tfrac32\ \ (HHH),\qquad
		2u+\tfrac12h\leq\tfrac32\ \ (HHL),\qquad
		u+v\leq\tfrac32\ \ (HLL),\qquad
		\tfrac32v\leq\tfrac32\ \ (LLL).
	\end{equation}

	Let $G_D=\sum_{z\in D}g(z)$. For the baselines, we verify that standard CRC with $B=1$ gives $1-\CRC(D)=1/(2G_D)$, and envelope CRC with the sharp envelope $U(\lambda)=1-\lambda$ gives $1-\ENV(D)=3/(2(G_D+1))$. In particular, envelope CRC has $(h,u,v)=(1/2,3/5,3/4)$.

	Starting from the controller of envelope CRC, the update \eqref{eq:recursion} at $D$ replaces the current coordinate $w_D$ by the largest $w\leq1$ with $g(z)(w-w_D)\leq\Delta_A(D\cup\{z\})/3$ for $z\in\{H,L\}$. The three coordinates evolve as follows:
	\begin{itemize}
		\item At $HH$, the candidate $H$ gives $w\leq h$, since $\Delta_A(HHH)=3/2-3h=0$ at $h=1/2$. Hence $h=1/2$ throughout.
		\item At $HL$, with $h=1/2$, the candidates $H$ and $L$ give $w\leq u/3+5/12$ and $w\leq u/3+1-2v/3$. The first bound is the smaller one whenever $v\leq7/8$.
		\item At $LL$, the candidates $H$ and $L$ give $w\leq 2v/3-u/3+1/2$ and $w\leq1$.
	\end{itemize}
	Hence, as long as $v_k\leq7/8$ and $2v_k/3-u_k/3+1/2\leq1$, the iterates satisfy
	\[
	u_{k+1}=\tfrac13u_k+\tfrac5{12},\qquad v_{k+1}=\tfrac23v_k-\tfrac13u_k+\tfrac12,
	\]
	with $u_0=3/5$ and $v_0=3/4$. By induction, the solutions are
	\[
	u_k=\frac58-\frac1{40}\left(\frac13\right)^k,\qquad
	v_k=\frac78-\frac1{10}\left(\frac23\right)^k-\frac1{40}\left(\frac13\right)^k.
	\]
	They satisfy $v_k<7/8$ for every $k$, so both conditions hold and the recurrences are the exact updates. In terms of actions,
	\[
	A^{(k)}(HH)=\frac12,\qquad
	A^{(k)}(HL)=\frac38+\frac1{40}\left(\frac13\right)^k,\qquad
	A^{(k)}(LL)=\frac18+\frac1{10}\left(\frac23\right)^k+\frac1{40}\left(\frac13\right)^k,
	\]
	which give the entries of Table~\ref{tab:toy} and converge geometrically at rate $\rho=2/3$ to $A^{(\infty)}=(1/2,3/8,1/8)$.

	In this case, a valid controller is admissible if and only if no other point satisfying \eqref{eq:toy-valid} is coordinatewise at least as large. We observe the following.
	\begin{itemize}
		\item The limit has $(h,u,v)=(1/2,5/8,7/8)$. The constraints for $HHH$, $HHL$ and $HLL$ bind, and the one for $LLL$ is slack, since $\frac32\cdot\frac78=\frac{21}{16}<\frac32$. If a feasible point $(h',u',v')$ is at least as large, then $HHH$ forces $h'=1/2$, $HHL$ then forces $u'=5/8$, and $HLL$ then forces $v'=7/8$. Hence the limit is admissible.
		\item The controller $(1/2,2/5,1/10)$ has $(h,u,v)=(1/2,3/5,9/10)$ and satisfies \eqref{eq:toy-valid}, with $HHH$ and $HLL$ binding. A feasible point at least as large has $h'=1/2$ by $HHH$, and $u'\geq3/5$, $v'\geq9/10$ and $u'+v'\leq3/2$ force $u'=3/5$ and $v'=9/10$. Hence this controller is also admissible.
		\item The two controllers are incomparable, and their pointwise minimum has $(h,u,v)=(1/2,5/8,9/10)$, which violates the constraint for $HLL$, since $5/8+9/10=61/40>3/2$.
	\end{itemize}

	Theorem~\ref{thm:crc-limit} requires $\Lambda$ to be an interval, even for affine losses. On the grid $\{1/10,1/4,2/5,1/2,1\}$, the envelope controller $(1/2,2/5,1/4)$ is a fixed point of the restricted update, since its feasible sets in \eqref{eq:recursion} at $HH$, $HL$ and $LL$, namely $[1/2,1/2]$, $[23/60,2/5]$ and $[1/5,1/4]$, contain no smaller grid value. Yet the valid grid controller $(1/2,2/5,1/10)$ strictly dominates it. The step that fails is the interpolation in Lemma~\ref{lem:one-step}: at $LL$, the action $\rho A(D)+C(D)/(n+1)=1/5$ is feasible but not on the grid. Nevertheless, the sweep of Proposition~\ref{prop:coordinate} needs no interpolation and reaches $(1/2,2/5,1/10)$ in one pass.

	\subsection{A two-risk example with a continuous observation space}\label{app:two-risk}
	We give one more example, with two risks and a continuous observation space, so that a controller is a function rather than a finite table. Recertification converges to the largest valid controller that dominates the joint envelope baseline, and the improvement is strict at every interior point.

	Take $n=1$ for simplicity, so that a single calibration observation $X\in[0,1]$ is used to choose an action for a test observation $Y\in[0,1]$, and let $\Lambda=[0,1]$. Throughout this example we describe actions by $a=1-\lambda$, so that larger values are less conservative, and write $A(x)\in[0,1]$ for the value chosen at the calibration observation $x$. The losses $ya$ and $(1-y)a$ must each have risk at most $1/4$,
	\[
	\E[YA(X)]\leq\tfrac14,\qquad \E[(1-Y)A(X)]\leq\tfrac14,
	\]
	under every exchangeable law of $(X,Y)$. Both losses are affine in $\lambda$, with sharp envelope $a$. By Lemma~\ref{lem:orbit}, applied to each risk, $A$ is jointly valid if and only if
	\begin{equation}\label{eq:two-risk-valid}
		yA(x)+xA(y)\leq\tfrac12\quad\text{and}\quad
		(1-y)A(x)+(1-x)A(y)\leq\tfrac12\qquad\text{for all }x,y\in[0,1].
	\end{equation}

	Envelope CRC for the two risks gives $A(x)\leq1/(2(1+x))$ and $A(x)\leq1/(2(2-x))$, and the joint envelope baseline $A_0$ takes the more conservative of the two. We compare it with $A_*$, where
	\[
	A_0(x)=\frac{1}{2(1+\max(x,1-x))},\qquad
	A_*(x)=\frac{1+\min(x,1-x)}4.
	\]
	If a jointly valid $A$ dominates $A_0$, then \eqref{eq:two-risk-valid} at $x=y=1$ and at $x=y=0$ forces $A(1)=A(0)=1/4$, the values of $A_0$ there, and the constraints with $y=1$ and $y=0$ then give $A(x)\leq\min\{(2-x)/4,(1+x)/4\}=A_*(x)$. Conversely, these two bounds give
	\[
	yA_*(x)+xA_*(y)\leq\frac{1-(1-x)(1-y)}2\leq\frac12,\qquad
	(1-y)A_*(x)+(1-x)A_*(y)\leq\frac{1-xy}2\leq\frac12,
	\]
	so $A_*$ is jointly valid. Hence $A_*$ is the largest valid controller that dominates $A_0$, and in particular it is admissible. The improvement is strict at every interior point.

	In terms of $a$, the update \eqref{eq:multi-recycle} at $x$ replaces $A(x)$ by the smallest of the thresholds
	\[
	\frac{A(x)}2+\frac{1/2-xA(y)}{2y}\quad(0<y\leq1),\qquad
	\frac{A(x)}2+\frac{1/2-(1-x)A(y)}{2(1-y)}\quad(0\leq y<1),
	\]
	which come from the first and the second risk. Suppose that $A\leq A_*$ and $A(0)=A(1)=1/4$, as for $A_0$. Then the tightest constraints come from the extreme test points, $y=1$ for the first risk and $y=0$ for the second. Indeed, for $0<y\leq1$ and $c\in[0,1]$,
	\[
	\frac12-cA(y)-y\left(\frac12-\frac c4\right)=\frac{1-y}2-c\left(A(y)-\frac y4\right)\geq\frac{1-y}2-c\left(A_*(y)-\frac y4\right)\geq0,
	\]
	since $A_*(y)-y/4=\min\{1/4,(1-y)/2\}\leq(1-y)/2$. With $c=x$, the first family of thresholds is therefore minimized at $y=1$, with value $A(x)/2+(2-x)/8$. Similarly, the second family is minimized at $y=0$, with value $A(x)/2+(1+x)/8$. The smaller of the two values is $\{A(x)+A_*(x)\}/2$, so
	\[
	A^{(k+1)}=\tfrac12\bigl(A^{(k)}+A_*\bigr).
	\]
	The updated controller again satisfies $A\leq A_*$ and $A(0)=A(1)=1/4$, and by induction
	\[
	A^{(k)}(x)=A_*(x)-2^{-k}\{A_*(x)-A_0(x)\}.
	\]
	The iterates thus increase to $A_*$, and the rate \eqref{eq:crc-rate} of Proposition~\ref{prop:multi-risk}, with $\rho=1/2$, holds with equality. Each iterate is a convex combination of $A_0$ and $A_*$, so its validity can also be read off from the linear constraints \eqref{eq:two-risk-valid}.

	\subsection{Implementation details for the real-data study}\label{app:simulation-implementation}
	All computations use the cached classifier outputs, and no model is fitted on the calibration or test images. Throughout, $A_k(D,x)$ denotes the depth assigned to a test image $x$ after $k$ rounds of recertification, starting from envelope CRC $A_0$; larger depths are less conservative.

	\paragraph{Error types.}
	For an observation $z=(x,y)$, let its type $c(z)$ be $0$ if $y=f(x)$, $1$ if $y\in g(x)\setminus\{f(x)\}$, and $2$ if $y\notin g(x)$. By \eqref{eq:simulation-hierarchy-loss}, the loss depends on $z$ only through its type,
	\[
	\ell_0(a)=0,\qquad \ell_1(a)=\tfrac12(a-1)_+,\qquad \ell_2(a)=\tfrac12a,\qquad a\in[0,2].
	\]
	The pool contains $7{,}300$, $969$ and $1{,}731$ images of the three types.

	\paragraph{Baselines.}
	\begin{itemize}
		\item \emph{Common depth.} With $h_j$ calibration observations of type $j$, CRC and envelope CRC select the largest $a\in[0,2]$ with
		\[
		\frac{h_1(a-1)_++h_2a}{2}+U(a)\le (n+1)\alpha,
		\qquad
		U(a)=1\ \text{(CRC)},\quad U(a)=a/2\ \text{(envelope CRC)},
		\]
		and the root, $a=0$, if no depth qualifies. The left side is piecewise linear with a kink at $a=1$, so the largest depth is found in closed form on $[0,1]$ and $[1,2]$. It is generally fractional, so these baselines already randomize between adjacent levels.
		\item \emph{Confidence-adaptive scores.} Write $u_i=1-p_g(X_i)$ and $v_i=1-p_f(X_i)$, so that $u_i\leq v_i$, and, under the threshold $t$, image $i$ moves from the root to its superclass when $t>u_i$ and to its leaf when $t>v_i$. The calibration loss at $t$ is
		\[
		\frac12\sum_{i=1}^n\left[\ind\{c(Z_i)=2\}\ind\{u_i<t\}+\ind\{c(Z_i)\ne0\}\ind\{v_i<t\}\right],
		\]
		a step function that increases by $1/2$ at the leaf breakpoint of every observation of nonzero type and at the superclass breakpoint of every observation of type $2$. CRC selects the largest $t$ at which at most $b_0=2((n+1)\alpha-1)$ of these jumps have occurred. These baselines return integer depths.
	\end{itemize}

	\paragraph{The update.}
	Fix a calibration sample $D$ and a test image $x$; in the common-depth family, the depth does not depend on $x$. For an augmented sample $S=(z_1,\ldots,z_{n+1})$, let $T_k(S)=\sum_{i=1}^{n+1}\ell(z_i,A_k(S_{-i},x_i))$ be the total held-out loss of $A_k$. Since the loss depends on the label only through its type, all labels of the same type give the same constraint in the update of Section~\ref{subsec:extensions}. For $c\in\{1,2\}$, let $S_c$ augment $D$ with $(x,y_c)$, where $y_c$ is any label of type $c$ for $x$, and let
	\[
	s_{k,c}(D,x)=\alpha-\frac{T_k(S_c)}{n+1}
	\]
	be the equal share of its unused budget. With $a=A_k(D,x)$, the next depth is the largest $a'\in[a,2]$ with $\ell_c(a')\leq\ell_c(a)+s_{k,c}(D,x)$ for $c=1,2$, that is,
	\begin{equation}\label{eq:appendix-update}
		A_{k+1}(D,x)=\min\left\{2,\ a+2s_{k,2}(D,x),\ 1+(a-1)_++2s_{k,1}(D,x)\right\}.
	\end{equation}

	For common depth, every iterate depends on $D$ only through the count vector $h=(h_0,h_1,h_2)$, so $A_k$ is a table with $\binom{n+2}{2}=1{,}326$ entries at $n=50$. For an augmented count vector $q$ with total $n+1$,
	\[
	T_k(q)=\sum_{j:\,q_j>0}q_j\,\ell_j\bigl(A_k(q-e_j)\bigr),
	\]
	where $e_j$ is the $j$th unit vector, and the slacks for the entry $h$ come from $q=h+e_1$ and $q=h+e_2$. Each round updates the whole table at once. The table does not depend on the data, so it is computed once for each target, and each split reads off the entry for its calibration counts.

	For confidence-adaptive scores, there is no common table, and for each test image we compute a table of all the depths that the recursion requires. Unrolling \eqref{eq:appendix-update}, $A_k(D,x)$ depends on depths $A_{k-1}(S_{-i},x_i)$, in which a calibration point is held out and the test image carries a hypothetical type, and these depend in turn on further depths of the same form. Every depth that arises is therefore indexed by a held-out point, among the calibration points and the test image, and a type for each remaining point. Calibration points of type $0$ never need to be held out, since their loss is zero, and they keep their type; they enter only through the sample size. If $r-1$ calibration points have nonzero types, call them and the test image the $r$ \emph{active} points. The table then holds one depth for each held-out active point and each assignment of type $1$ or $2$ to the other active points, giving $r2^{r-1}$ entries in total. Each round updates all entries from the preceding table, in time linear in its size, and the reported depth is the entry in which the test image is held out and the calibration points carry their observed types. Consequently, the computation can be more demanding than in the common-depth case. Two exact reductions shrink the table without changing any depth.
	\begin{itemize}
		\item \emph{Capping breakpoints.} Let $b=\lfloor b_0\rfloor$ be the number of whole jumps within the budget. Every active point has a leaf breakpoint, whatever its type. If there are at least $b+2$ active points, let $\tau$ be the $(b+2)$th smallest of their leaf breakpoints. After any one point is held out, at least $b+1$ jumps at or below $\tau$ remain, so every CRC threshold in the table is at most $\tau$. Replacing each breakpoint above $\tau$ by $\tau$ therefore leaves all baseline depths unchanged, and hence all later ones.
		\item \emph{Grouping.} Active points with identical breakpoint pairs, after capping, are interchangeable, so a group of $g$ such points is described by its number of type-$2$ points rather than by $2^g$ assignments. With group sizes $g_1,\ldots,g_J$, the table has $\sum_jg_j\prod_{h\ne j}(g_h+1)$ entries.
	\end{itemize}
	A larger nominal level allows more jumps, which raises $\tau$ and leaves more distinct breakpoints. This explains the longer runtimes at larger levels.

	We record rounds $0,1,3,10,20,50,100,200$ and $500$, and the common-depth table is updated for all $500$ rounds. For an adaptive test image, the total slack $(n+1)\alpha-T_k(S_c)$ bounds how much its loss can grow in all later rounds. Once this slack allows a depth increase of at most $10^{-8}$, we stop updating the image and carry its depth forward, which preserves validity by the first observation at the end of Section~\ref{subsec:extensions}.

	\paragraph{Evaluation.}
	Each of the $300$ splits draws $50$ calibration images and $16$ test images at random without replacement from the pool of $10{,}000$ images, and all methods, families and targets use the same splits. All metrics are computed exactly over the randomization of the prediction. With $\pi(a)=((1-a)_+,1-|a-1|,(a-1)_+)$ the probabilities of the root, superclass and leaf, a test image of type $c$ predicted at depth $a$ has risk $\ell_c(a)$, expected node size $100\pi_0(a)+5\pi_1(a)+\pi_2(a)$ and expected correct specificity
	\[
	\frac{\log20}{\log100}\,\pi_1(a)\ind\{c\ne2\}+\pi_2(a)\ind\{c=0\}.
	\]
	Each metric is averaged over the $16$ test images of a split and then over the splits, and standard errors are computed across splits. In Figure~\ref{fig:simulation-trajectory}, let $d_s$ be the gain over envelope CRC in split $s$, and let $\bar d$ and $\mathrm{se}$ be the mean and standard error of $d_1,\ldots,d_M$ with $M=300$. The bands are $\bar d\pm t_{M-1,0.975}\,\mathrm{se}$, pointwise intervals that reflect variation across splits of the fixed pool.

	\section{Details for randomized e-BH}\label{app:rebh}
	Throughout, $e$ is a fixed nonnegative vector, $b_{[r]}=(\alpha/K)e_{[r]}$ is the budget of the hypothesis of rank $r$, and $c_{[r]}(p)=\sum_{k\geq r}p_k/k$ is its charge under a distribution $p$ of the number of rejections, as in Section~\ref{subsec:rebh-slack}. We also write $a_r=r\,b_{[r]}+\sum_{j>r}b_{[j]}$, so that $q^*_r=\min\{1,a_r\}$.

	\subsection{The certified menu and U-eBH}\label{app:rebh-menu}
	We first check that e-BH is the largest certified set.
	By Section~\ref{subsec:rebh}, the sets certified by $E^{\mathrm{eBH}}$ are the self-consistent ones. Let $k^*$ be the e-BH count, the largest $k$ with $k\,e_{[k]}\geq K/\alpha$. The e-BH set is self-consistent, because each of its members has an e-value of at least $e_{[k^*]}\geq K/(\alpha k^*)$. Moreover, recall that e-BH rejects every hypothesis $j$ with $e_j\geq K/(\alpha k^*)$. If $R$ is self-consistent and $|R|=k$, then its $k$ members have e-values of at least $K/(\alpha k)$, so $k\,e_{[k]}\geq K/\alpha$ and $k\leq k^*$. Every member of $R$ then has $e_j\geq K/(\alpha k)\geq K/(\alpha k^*)$ and is rejected by e-BH.

	We also verify explicitly that every boosting rule that selects from the certified menu \eqref{eq:rebh-budget} controls the FDR.
	If $p(e)$ is measurable in $e$ and lies in the menu for every $e$, then for $P\in\Pc_T$, since the number of rejections is drawn independently of the data,
	\[
	\E\,\FDP_T(R_\kappa(E))=\E_P\bar L(T,p(E);E)=\E_P\sum_{j\in T}c_{[r(j)]}(p(E))\leq\frac\alpha K\sum_{j\in T}\E_PE_j\leq\frac{\alpha|T|}K .
	\]
	This is Theorem~\ref{thm:closure} for the action space of distributions.

	As described in the main text, the rule given by U-eBH leaves a nonnegative unused budget.
	Fix $r$, and let $k_r\geq r$ maximize $k\,e_{[k]}$ over $k\geq r$. By \eqref{eq:uebh-tails}, the tails of $p^0$ are equal at all ranks $r'\in\{r,\ldots,k_r\}$, since the maximum of $k\,e_{[k]}$ over $k\geq r'$ is still attained at $k_r$. Since $\{\kappa_0\geq k\}=\{U\leq\Prob(\kappa_0\geq k\mid E=e)\}$, the events $\{\kappa_0 \geq r\}$ and $\{\kappa_0\geq k_r\}$ coincide. Whenever U-eBH rejects the hypothesis of rank $r$, it rejects at least $k_r$ hypotheses, so that hypothesis receives a share of at most $1/k_r$. Hence
	\[
	c_{[r]}(p^0)=\E\left[\frac{\ind\{\kappa_0\geq r\}}{\kappa_0}\Bigm|E=e\right]\leq\frac{\Prob(\kappa_0\geq r\mid E=e)}{k_r}\leq\frac\alpha K\,e_{[k_r]}\leq b_{[r]} .
	\]
	U-eBH therefore selects from the menu, and the difference $b_{[r]}-c_{[r]}(p^0)$ is the budget that it leaves unused for the hypothesis of rank $r$, which is positive whenever $e_{[r]}>e_{[k_r]}$.

	Recertifying U-eBH with $E^{\mathrm{eBH}}$ is exactly an instance of Proposition~\ref{prop:conditional}. The state is the set $T$ of true nulls, and the loss is the averaged false discovery proportion. Since U-eBH selects the single distribution $p^0(e)$, its canonical envelope is its own loss, $g_T=\bar L(T,p^0(E);E)=\sum_{j\in T}c_{[r(j)]}(p^0(E))$. Take the conditioning statistic $V_T=E$. Given $E$, the loss $g_T$ is known, because the law of $U$ is known, so we may take $H_T=g_T$. Take the budget $B_T=\alpha E^{\mathrm{eBH}}_T=(\alpha/K)\sum_{j\in T}E_j$. Summing the previous paragraph over $j\in T$ gives $H_T\leq B_T$, and $\E_PB_T\leq\alpha|T|/K\leq\alpha$ for $P\in\Pc_T$, so \eqref{eq:conditional-budget} holds. The allowance $B_T-H_T$ is the unused budget of the previous paragraph, and it is a function of $E$. The increment $W_T=B_T-H_T$ therefore satisfies \eqref{eq:conditional-increment} with equality, and the recertified e-collection $(g_T+W_T)/\alpha=B_T/\alpha$ is $E^{\mathrm{eBH}}_T$. Its certified menu is \eqref{eq:rebh-budget}, which contains $p^0$ and every other distribution whose charges stay within the budgets.

	\subsection{Proof of Proposition~\ref{prop:rebh}}\label{app:rebh-proof}
	Section~\ref{subsec:rebh-recert} shows that every distribution in the menu \eqref{eq:rebh-budget} satisfies the tail bounds \eqref{eq:rebh-tails}. The following lemma shows that a single distribution in the menu attains all of them. Put $e_{[K+1]}=0$ and $a_{K+1}=0$ for notational convenience.

	\begin{lemma}\label{lem:rebh-attain}
		Let $\tilde p_k=(\alpha k/K)(e_{[k]}-e_{[k+1]})$ for $k=1,\ldots,K$, define $p^*_k=\min\{\tilde p_k,1-\sum_{l>k}p^*_l\}$ for $k=K,K-1,\ldots,1$, and let $p^*_0=1-\sum_{k\geq1}p^*_k$. Then $p^*$ is a distribution in the menu \eqref{eq:rebh-budget} with tails $\sum_{k\geq r}p^*_k=q^*_r$ for $r=1,\ldots,K$. If $a_1\leq1$, then $p^*_k=\tilde p_k$ for $k\geq1$, and every constraint in \eqref{eq:rebh-budget} holds with equality.
	\end{lemma}
	\begin{proof}
		We check that $\tilde p_k$ is nonnegative, $\sum_{k\geq r}\tilde p_k/k=(\alpha/K)\sum_{k\geq r}(e_{[k]}-e_{[k+1]})=b_{[r]}$, and $\tilde p_r=a_r-a_{r+1}$. By construction, $0\leq p^*_k\leq\tilde p_k$ for $k\geq1$ and $\sum_{k\geq1}p^*_k\leq1$, so $p^*$ is a distribution whose charges are at most the budgets $b_{[r]}$, and it lies in the menu. Downward induction in $r$ gives
		\[
		\sum_{k\geq r}p^*_k=\min\Bigl\{1,\sum_{k>r}p^*_k+\tilde p_r\Bigr\}=\min\bigl\{1,\min\{1,a_{r+1}\}+\tilde p_r\bigr\}=\min\{1,a_r\}=q^*_r .
		\]
		If $a_1\leq1$, then $\tilde p_k\leq1-a_{k+1}$ for every $k$, so no truncation occurs, $p^*_k=\tilde p_k$ for $k\geq1$, and every charge equals its budget.
	\end{proof}

	\begin{proof}[Proof of Proposition~\ref{prop:rebh}]
		By Lemma~\ref{lem:rebh-attain}, the bounds $q^*_r$ are the tails of the distribution $p^*$, so they are non-increasing in $r$. Hence $\{\kappa^*\geq r\}=\{U\leq q^*_r\}$, and $\kappa^*$ has the distribution $p^*$ given $E=e$.

		(i) The tails $q^*_r(e)$ are measurable in $e$, so $p^*(e)$ is a measurable selection from the menu \eqref{eq:rebh-budget}, and the validity of boosting rules in Appendix~\ref{app:rebh-menu} gives $\E\,\FDP_T(R_{\kappa^*}(E))\leq\alpha|T|/K$.

		(ii) If $k\geq r$, then
		\[
		r\,e_{[r]}+\sum_{j>r}e_{[j]}\geq r\,e_{[k]}+\sum_{j=r+1}^ke_{[j]}\geq k\,e_{[k]},
		\]
		so $q^*_r$ is at least the U-eBH tail \eqref{eq:uebh-tails}. With the same $U$, it follows that $\{\kappa_0\geq r\}\subseteq\{\kappa^*\geq r\}$ for every $r$, so $\kappa^*\geq\kappa_0$ and $R_{\kappa_0}(e)\subseteq R_{\kappa^*}(e)$. Clearly, U-eBH in turn contains e-BH.

		(iii) Every boosting rule that selects from \eqref{eq:rebh-budget} has tails at most $q^*_r$ by \eqref{eq:rebh-tails}, and $\kappa^*$ attains them.
	\end{proof}

	\begin{corollary}\label{cor:rebh-evalues}
		Let $\widetilde E_j=(K/\alpha)\ind\{j\in R_{\kappa^*}\}/(|R_{\kappa^*}|\vee1)$. Then $\E[\widetilde E_j\mid E]\leq E_j$, so $\widetilde E_j$ is an e-value whenever $E_j$ is, and e-BH applied to $\widetilde E$ rejects exactly $R_{\kappa^*}(E)$.
	\end{corollary}
	\begin{proof}
		Given $E=e$, $\E[\widetilde E_j\mid E=e]=(K/\alpha)\,c_{[r(j)]}(p^*)\leq(K/\alpha)\,b_{[r(j)]}=e_j$. Every member of $R_{\kappa^*}$ has $\widetilde E_j=K/(\alpha|R_{\kappa^*}|)$ and every other hypothesis has $\widetilde E_j=0$, so the largest $k$ with $k\,\widetilde E_{[k]}\geq K/\alpha$ is $|R_{\kappa^*}|$, and e-BH applied to $\widetilde E$ rejects exactly $R_{\kappa^*}$.
	\end{proof}

	\subsection{Randomization over arbitrary rejection sets}\label{app:rebh-general}
	The main text considers randomized rules that reject one of the nested sets $R_k(e)$. More generally, a randomized rule may choose, given $E=e$, any distribution $Q$ over the subsets $R\subseteq\{1,\ldots,K\}$. Each rejected null hypothesis contributes $1/|R|$ to the false discovery proportion, so in state $T$,
	\[
	\E_Q\FDP_T(R)=\sum_{j\in T}c_j(Q),\qquad c_j(Q)=\E_Q\biggl[\frac{\ind\{j\in R\}}{|R|\vee1}\biggr],
	\]
	where $c_j(Q)$ is the expected share charged to hypothesis $j$. As in Section~\ref{subsec:rebh}, both sides of $\E_Q\FDP_T(R)\leq\alpha E^{\mathrm{eBH}}_T(e)$ are sums over $T$, so $Q$ is certified by $E^{\mathrm{eBH}}$ if and only if $c_j(Q)\leq b_j=(\alpha/K)e_j$ for every $j$.

	J-eBH is such a rule whose rejection sets are not nested, and it is certified by $E^{\mathrm{eBH}}$, so that in particular it controls the FDR. To bound the charge of hypothesis $j$, fix $E=e$ and the uniforms $U_i$ for $i\neq j$, and let $f(u)$ be the number of rejections when $U_j=u$. Increasing $u$ decreases the boosted value $e_j/u$ and leaves the others unchanged, so $f$ is non-increasing. Since e-BH rejects every hypothesis whose boosted value is at least $K/(\alpha f(u))$, hypothesis $j$ is rejected exactly when $u\leq(\alpha/K)e_jf(u)$. If this condition holds at $u$, it holds at every $u'<u$, because $u'<u\leq(\alpha/K)e_jf(u)\leq(\alpha/K)e_jf(u')$. The values of $u$ at which hypothesis $j$ is rejected therefore form an interval with left endpoint $0$ and some right endpoint $u^*$. If $u^*>0$, then $f(u)\geq f^-(u^*)\geq1$ on this interval, where $f^-(u^*)$ is the left limit of $f$ at $u^*$, and letting $u\uparrow u^*$ in the rejection condition gives $u^*\leq(\alpha/K)e_jf^-(u^*)$. Hence
	\[
	\E\biggl[\frac{\ind\{j\in R\}}{|R|\vee1}\biggm|U_i,\,i\neq j\biggr]=\int_0^{u^*}\frac{du}{f(u)}\leq\frac{u^*}{f^-(u^*)}\leq\frac\alpha Ke_j=b_j,
	\]
	which also holds trivially if $u^*=0$, and averaging over the other uniforms gives $c_j(Q)\leq b_j$.

	\begin{proposition}\label{prop:rebh-general}
		Every distribution $Q$ over rejection sets that is certified by $E^{\mathrm{eBH}}$ satisfies $\E_Q|R|\leq\E[\kappa^*\mid E=e]$. The stochastic comparison of Proposition~\ref{prop:rebh}(iii) does not extend to this class: such a $Q$ may satisfy $Q(|R|\geq r)>q^*_r$ for some $r$.
	\end{proposition}
	\begin{proof}
		Let $m$ be the number of ranks $r$ with $a_r\geq1$. Since $a_r$ is non-increasing in $r$, $q^*_r=1$ for $r\leq m$ and $q^*_r=a_r$ for $r>m$. Summing the tails,
		\[
		\E[\kappa^*\mid E=e]=\sum_{r=1}^Kq^*_r=m+\sum_{r>m}a_r=m+\sum_{j>m}(2j-m-1)\,b_{[j]},
		\]
		because for $j>m$, the budget $b_{[j]}$ enters $a_j$ with coefficient $j$ and $a_r$ with coefficient one for each $m<r<j$. Give the hypothesis $j$ the weight $y_j=0$ if $r(j)\leq m$ and $y_j=2r(j)-m-1$ otherwise, so that $\E[\kappa^*\mid E=e]=m+\sum_jy_jb_j$. The weights, ordered by rank, are $m$ zeros followed by $m+1,m+3,m+5,\ldots$, so the weights in a set $R$ of size $k\geq1$ sum to at least $0$ if $k\leq m$ and at least $\sum_{i=1}^{k-m}(m+2i-1)=k(k-m)$ otherwise. In both cases $\sum_{j\in R}y_j\geq k(k-m)$, which leads to
		\[
		|R|\leq m\,\ind\{R\neq\emptyset\}+\sum_jy_j\frac{\ind\{j\in R\}}{|R|\vee1}.
		\]
		This also holds for $R=\emptyset$. Taking expectations under $Q$ and using $c_j(Q)\leq b_j$,
		\[
		\E_Q|R|\leq m\,Q(R\neq\emptyset)+\sum_jy_j\,c_j(Q)\leq m+\sum_jy_jb_j=\E[\kappa^*\mid E=e].
		\]
		The constant $m$ and the weights $y_j$ thus act as a dual certificate for the linear program that maximizes $\E_Q|R|$ subject to the budget constraints.

		For the second claim, take $K=5$, $\alpha=0.05$ and $e=(100,1,1,1,1)$, so that $b_1=1$ and $b_j=0.01$ for $j\geq2$. Let $Q$ always reject $H_1$ and reject each other hypothesis independently with probability $0.02$. Then $c_1(Q)\leq1=b_1$, and for $j\geq2$, every rejection of $H_j$ comes with that of $H_1$, so $c_j(Q)\leq0.02/2=b_j$. Hence $Q$ is certified, yet $Q(|R|\geq2)=1-0.98^4\approx0.078$, whereas $q^*_2=0.05$. Its expected number of rejections, $1.08$, is below $\E[\kappa^*\mid E=e]=1.2$, as the first claim requires.
	\end{proof}

	\subsection{Proof of Proposition~\ref{prop:rebh-mean}}\label{app:rebh-mean}

	Throughout, $e$ is fixed, $S=S^\sharp$, and $c_r=S_r-S_{r+1}$ for $r=1,\ldots,K$, with $c_{K+1}=0$. We show below that these are the charges of the rule. A state is identified with the set of ranks of its hypotheses, so that $\{s,\ldots,K\}$ denotes the $K-s+1$ hypotheses with the smallest e-values. Write $\Phi_s=\alpha\sum_{j\geq s}e_{[j]}$ for $s=1,\ldots,K+1$, so that $\alpha e_{[r]}=\Phi_r-\Phi_{r+1}$ and $\Phi_s=(K-s+1)\alpha m_s$. The sequence $\Phi$ is convex, because its decrements $\alpha e_{[s]}$ are non-increasing. We call $s$ a vertex if $S_s=\beta_s$. The points $1$ and $K+1$ are vertices, and $S$ is linear between consecutive vertices. Since $S\leq\beta$ and $\beta_s\leq\alpha m_s$, we have $(K-s+1)S_s\leq\Phi_s$ for every $s$, with equality at every vertex $s\geq2$. We first prove three lemmas that are central to the proof of Proposition~\ref{prop:rebh-mean}.

	\begin{lemma}\label{lem:mean-rule}
		The sequence $S$ is nonnegative, non-increasing and convex, and the charges $c_r$ are nonnegative and non-increasing in $r$. The numbers $p^\sharp_k=k(c_k-c_{k+1})$ for $k=1,\ldots,K$ and $p^\sharp_0=1-S_1$ form a distribution with charges $c_{[r]}(p^\sharp)=c_r$ and tails $q^\sharp_r$, and $\kappa^\sharp$ has the distribution $p^\sharp$ given $E=e$. Moreover, $q^\sharp_r$ is a continuous function of $e$.
	\end{lemma}
	\begin{proof}
		The zero sequence is a convex minorant of $\beta$, so $S\geq0$, and a nonnegative convex sequence that vanishes at $K+1$ is non-increasing. Hence $c_r\geq0$, and $c_r$ is non-increasing in $r$ by convexity, so $p^\sharp_k\geq0$ for $k\geq1$. Since $\sum_{k\geq1}p^\sharp_k=\sum_{k\geq1}c_k=S_1\leq\beta_1\leq1$, also $p^\sharp_0\geq0$, and the numbers sum to one. Telescoping gives $\sum_{k\geq r}p^\sharp_k/k=\sum_{k\geq r}(c_k-c_{k+1})=c_r$, so $p^\sharp$ has the charges $c_r$, and by the accounting identity of Section~\ref{subsec:rebh-recert} its tails are $r\,c_r+\sum_{j>r}c_j=rS_r-(r-1)S_{r+1}=q^\sharp_r$. Tails of a distribution are non-increasing, so $\{\kappa^\sharp\geq r\}=\{U\leq q^\sharp_r\}$, and $\kappa^\sharp$ has the distribution $p^\sharp$. Finally, $\beta$ is continuous in $e$, and so is its greatest convex minorant.
	\end{proof}

	\begin{lemma}\label{lem:mean-below}
		If a distribution $p$ of the number of rejections satisfies $S_s(p)\leq\alpha m_s$ for $s=1,\ldots,K$, then $S(p)\leq S$. In particular, this holds for every monotone rule certified by $E^{\overline{\mathrm{eBH}}}$.
	\end{lemma}
	\begin{proof}
		Since also $S_1(p)=\sum_{k\geq1}p_k\leq1$, we have $S(p)\leq\beta$. The charges $c_{[r]}(p)$ are non-increasing in $r$, so $S(p)$ is convex with $S_{K+1}(p)=0$, and it is therefore a convex minorant of $\beta$, which lies below the greatest one, $S$. The constraints $S_s(p)\leq\alpha m_s$ are the constraints \eqref{eq:mean-menu} for the states $\{s,\ldots,K\}$.
	\end{proof}

	\begin{lemma}\label{lem:mean-exchange}
		For $1\leq r<r'\leq K$,
		\begin{equation}\label{eq:mean-exchange}
			(K-r)(c_r-c_{r'})\leq\alpha(e_{[r]}-e_{[r']}).
		\end{equation}
	\end{lemma}
	\begin{proof}
		Let $w$ be the smallest vertex greater than $r$, and write $\gamma=c_r$, the slope magnitude of the linear piece of $S$ that ends at $w$. If $r'<w$, then $c_{r'}=\gamma$, and \eqref{eq:mean-exchange} holds because $e_{[r]}\geq e_{[r']}$. Otherwise, let $v$ be the largest vertex with $v\leq r'$, so that $2\leq w\leq v$, $c_{r'}=c_v=:\gamma'$ and $e_{[r']}\leq e_{[v]}$. We bound $e_{[r]}$ from below and $e_{[v]}$ from above. Since $\Phi$ is convex and $r<w$, its decrement at $r$ is at least its average decrement over $[r,w]$. With $\Phi_r\geq(K-r+1)S_r$, $\Phi_w=(K-w+1)S_w$ and $S_r=S_w+(w-r)\gamma$, this gives
		\[
		\alpha e_{[r]}\geq\frac{\Phi_r-\Phi_w}{w-r}\geq\frac{(K-r+1)S_r-(K-w+1)S_w}{w-r}=(K-r+1)\gamma+S_w.
		\]
		With $\Phi_v=(K-v+1)S_v$ and $\Phi_{v+1}\geq(K-v)S_{v+1}$,
		\[
		\alpha e_{[v]}=\Phi_v-\Phi_{v+1}\leq(K-v+1)S_v-(K-v)S_{v+1}=S_v+(K-v)\gamma'.
		\]
		Since the slopes are non-increasing, $S_w-S_v=\sum_{i=w}^{v-1}c_i\geq(v-w)\gamma'$. Combining the three bounds,
		\[
		\alpha(e_{[r]}-e_{[r']})-(K-r)(\gamma-\gamma')\geq(K-r+1)\gamma+(v-w)\gamma'-(K-v)\gamma'-(K-r)(\gamma-\gamma')=\gamma+(2v-w-r)\gamma'\geq0,
		\]
		since $v\geq w>r$.
	\end{proof}

	\begin{proof}[Proof of Proposition~\ref{prop:rebh-mean}]
		(i) Fix $t$ and a state $T$ with $|T|=t$, let $U_t=\{K-t+1,\ldots,K\}$, and let $F(T)=\sum_{r\in T}(t\,c_r-\alpha e_{[r]})$, so that, by Lemma~\ref{lem:mean-rule}, the constraint \eqref{eq:mean-menu} for $T$ reads $F(T)\leq0$. If $T\neq U_t$, choose $r\in T\setminus U_t$ and $r'\in U_t\setminus T$. Then $r\leq K-t<r'$, so $t\leq K-r$, and replacing $r$ by $r'$ in $T$ changes $F$ by
		\[
		\alpha(e_{[r]}-e_{[r']})-t(c_r-c_{r'})\geq\alpha(e_{[r]}-e_{[r']})-(K-r)(c_r-c_{r'})\geq0,
		\]
		by $c_r\geq c_{r'}$ and Lemma~\ref{lem:mean-exchange}. Repeating such exchanges reaches $U_t$, so
		\[
		F(T)\leq F(U_t)=t\,(S_{K-t+1}-\alpha m_{K-t+1})\leq0.
		\]
		Every constraint in \eqref{eq:mean-menu} therefore holds. By Lemma~\ref{lem:mean-rule}, $p^\sharp(e)$ is a measurable selection from the certified menu of $E^{\overline{\mathrm{eBH}}}$, and Theorem~\ref{thm:closure} gives $\E\,\FDP_T(R_{\kappa^\sharp}(E))\leq\alpha\,\E_PE^{\overline{\mathrm{eBH}}}_T\leq\alpha$ for every nonempty $T$ and $P\in\Pc_T$.

		(ii) Fix $r$, and let $v<w$ be consecutive vertices with $v\leq r<w$, so that $c_r=c_v$ and $S_r=S_v-(r-v)c_v$. Then
		\[
		q^\sharp_r=rS_r-(r-1)S_{r+1}=S_v+(v-1)c_v.
		\]
		If $v=1$, then $q^\sharp_r=S_1=\min\{1,\alpha m_1\}=q^*_1\geq q^*_r$. If $v=K$, then $r=K$, $c_K=S_K=\alpha e_{[K]}$ and $q^\sharp_K=K\alpha e_{[K]}\geq q^*_K$. Otherwise $2\leq v\leq K-1$, so $S_v=\alpha m_v$ and $c_v=S_v-S_{v+1}\geq\alpha(m_v-m_{v+1})$. With $d=K-v+1$ and $\bar e=m_{v+1}$, we have $m_v-m_{v+1}=(e_{[v]}-\bar e)/d$ and $(d-1)\bar e=\sum_{j>v}e_{[j]}$, and a direct computation gives
		\[
		q^\sharp_r\geq\alpha m_v+(v-1)\,\alpha\,\frac{e_{[v]}-\bar e}{d}
		=\frac\alpha K\Bigl(v\,e_{[v]}+\sum_{j>v}e_{[j]}\Bigr)+\frac{\alpha\,v(v-1)(e_{[v]}-\bar e)}{dK}\geq q^*_v\geq q^*_r,
		\]
		because $e_{[v]}\geq\bar e$ and $q^*$ is non-increasing. In all cases $q^\sharp_r\geq q^*_r$, so with the same $U$, $\{\kappa^*\geq r\}\subseteq\{\kappa^\sharp\geq r\}$ for every $r$, and $R_{\kappa^*}(e)\subseteq R_{\kappa^\sharp}(e)$.

		(iii) Consider the linear program that maximizes $\sum_{k=1}^Kk\,p_k$ over $p_1,\ldots,p_K\geq0$ with $\sum_kp_k\leq1$, subject only to the constraints for the states $\{s,\ldots,K\}$,
		\[
		S_s(p)=\sum_{k=1}^Kp_k\frac{(k-s+1)_+}{k}\leq\alpha m_s,\qquad s=1,\ldots,K.
		\]
		By the accounting identity, its objective is $\sum_r(2r-1)c_{[r]}(p)=S_1(p)+2\sum_{s\geq2}S_s(p)$, which increases with every $S_s(p)$. Every feasible $p$ satisfies $S(p)\leq S$ by Lemma~\ref{lem:mean-below}, and $p^\sharp$ is feasible with $S(p^\sharp)=S$, so the value of the program is $\E[\kappa^\sharp\mid E=e]$. By linear programming duality, there are $\lambda_1,\ldots,\lambda_K\geq0$ and $\mu\geq0$ with $\sum_s\lambda_s\alpha m_s+\mu=\E[\kappa^\sharp\mid E=e]$ and
		\[
		\sum_{s=1}^K\lambda_s\frac{(k-s+1)_+}{k}+\mu\geq k,\qquad k=1,\ldots,K.
		\]
		A set $R$ of size $k\geq1$ contains at least $(k-s+1)_+$ of the $K-s+1$ hypotheses of rank at least $s$, so
		\[
		|R|\leq\sum_{s=1}^K\lambda_s\frac{|R\cap\{s,\ldots,K\}|}{|R|\vee1}+\mu\,\ind\{R\neq\varnothing\},
		\]
		which also holds for $R=\varnothing$. Let $Q$ be a distribution of rejection sets certified by $E^{\overline{\mathrm{eBH}}}$. Its averaged false discovery proportion in the state $\{s,\ldots,K\}$ is at most $\alpha m_s$, so taking expectations under $Q$ gives
		\[
		\E_Q|R|\leq\sum_s\lambda_s\,\E_Q\Bigl[\frac{|R\cap\{s,\ldots,K\}|}{|R|\vee1}\Bigr]+\mu\,Q(R\neq\varnothing)\leq\sum_s\lambda_s\alpha m_s+\mu=\E[\kappa^\sharp\mid E=e].
		\]
		The rejection set of closed e-BH is certified by $E^{\overline{\mathrm{eBH}}}$, which proves the last assertion.
	\end{proof}

	\subsection{The minimally adaptive improvement}\label{app:rebh-ma}
	We show that the minimally adaptive principle yields a valid rule that rejects at least as much as RU-eBH and sometimes strictly more, so that RU-eBH is not admissible. For a level $\gamma>0$, write $\kappa^*_\gamma$ for the rule $\kappa^*$ computed at level $\gamma$ instead of $\alpha$, with the same $U$, so that $\kappa^*=\kappa^*_\alpha$. Let $K\geq2$ and $\alpha'=K\alpha/(K-1)$, and define
	\[
	R^\dagger=\begin{cases}R_{\kappa^*_{\alpha'}}(E),&\kappa^*_\alpha>0,\\ \emptyset,&\kappa^*_\alpha=0.\end{cases}
	\]

	\begin{proposition}\label{prop:rebh-ma}
		The rule $R^\dagger$ has the following properties.
		\begin{enumerate}[label=(\roman*),leftmargin=*,itemsep=0pt]
			\item Its FDR is at most $\alpha$ in every state $T$ and under every law in $\Pc_T$.
			\item It contains the RU-eBH rejection set for every $E$ and $U$, and under some law it makes strictly more true discoveries with positive probability.
			\item It is certified by $E^{\overline{\mathrm{eBH}}}$, so RC-eBH rejects at least as many hypotheses as $R^\dagger$ in expectation given $E=e$.
		\end{enumerate}
	\end{proposition}
	\begin{proof}
		Let $B^\dagger_T=E^{\mathrm{eBH}}_T$ for the full set $T=\{1,\ldots,K\}$ and $B^\dagger_T=(K-1)^{-1}\sum_{j\in T}E_j$ for every other nonempty $T$. This is an e-collection, because a state other than the full set contains at most $K-1$ true nulls, and it exceeds $E^{\mathrm{eBH}}$ on every proper nonempty state.

		(i) Given $E=e$, $R^\dagger$ is a monotone rule, and we bound its averaged loss in each state. In the full state, every rejection is false, so the averaged loss is $\Prob(\kappa^*_\alpha>0\mid E=e)=q^*_1\leq(\alpha/K)\sum_je_j=\alpha B^\dagger_T$. In any other nonempty state $T$, the set $R^\dagger$ is either $R_{\kappa^*_{\alpha'}}$ or empty, so its averaged loss is at most that of $R_{\kappa^*_{\alpha'}}$. By the menu \eqref{eq:rebh-budget} at level $\alpha'$, the latter is at most $(\alpha'/K)\sum_{j\in T}e_j=\alpha B^\dagger_T$. Hence $R^\dagger$ selects from the certified menu of $B^\dagger$, measurably in $e$, and Theorem~\ref{thm:closure} gives an FDR of at most $\alpha\,\E_PB^\dagger_T\leq\alpha$.

		(ii) The tails $q^*_r$ are non-decreasing in the level, so $\kappa^*_{\alpha'}\geq\kappa^*_\alpha$, and $R^\dagger\supseteq R_{\kappa^*_\alpha}(E)$ for every $E$ and $U$. For the strict improvement, let $e^\circ=(1/\alpha,1/(2\alpha),\ldots,1/(2\alpha))$. At level $\alpha$, $q^*_1=(K+1)/(2K)$ and $q^*_r=1/2$ for $r\geq2$, whereas at level $\alpha'$, every $q^*_r$ is at least $K/(2(K-1))>(K+1)/(2K)$. For $1/2<U\leq(K+1)/(2K)$, RU-eBH therefore rejects only $H_1$, whereas $R^\dagger$ rejects all $K$ hypotheses. Under the law with $T=\{1\}$ in which $E=e^\circ$ with probability $\alpha$ and $E=0$ otherwise, $R^\dagger$ makes $K-1$ more true discoveries than RU-eBH with probability $\alpha/(2K)$.

		(iii) Since a nonempty state $T\neq\{1,\ldots,K\}$ has $|T|\leq K-1$, we have $B^\dagger\leq E^{\overline{\mathrm{eBH}}}$, with equality on the full set. The certified menu of $B^\dagger$ is therefore contained in that of $E^{\overline{\mathrm{eBH}}}$, and Proposition~\ref{prop:rebh-mean}(iii) applies.
	\end{proof}

	\subsection{Conditional calibration as recertification}\label{app:cc}
	This subsection spells out the correspondence summarized in Remark~\ref{rem:cc}. We restate the e-BH version of conditional calibration of \citet{leeren2024} in our terms, without modifying it. Let $e_1(X),\ldots,e_K(X)$ be finite e-values computed from data $X$, let $R_0=\eBH_\alpha(e(X))$ and $r=|R_0|$, and for each $j$ and $c\geq1$ let
	\[
	t_j(X)=\frac{K}{\alpha\,|R_0\cup\{j\}|},\qquad
	h_j(c,X)=t_j(X)\ind\{c\,e_j(X)\geq t_j(X)\}.
	\]
	For $j\in R_0$, $t_j=K/(\alpha r)$ is the self-consistency threshold that $e_j$ already passes, and for $j\notin R_0$, $t_j=K/(\alpha(r+1))$ is the value that $e_j$ would need for $R_0\cup\{j\}$ to be self-consistent. The quantity $h_j(c,X)$ inflates $e_j$ by the factor $c$ and then records only whether the threshold is reached; it is non-decreasing in $c$.

	At $c=1$, these are the pruned e-values of Remark~\ref{rem:cc},
	\[
	h_j(1,X)=\frac{K\ind\{j\in R_0\}}{\alpha(r\vee1)}\leq e_j(X).
	\]
	Indeed, every member of $R_0$ passes its threshold, whereas no $j\notin R_0$ can satisfy $e_j\geq K/(\alpha(r+1))$, since all $r+1$ members of $R_0\cup\{j\}$ would then reach this value and e-BH would reject at least $r+1$ hypotheses. Applying e-BH to the pruned e-values therefore rejects exactly $R_0$. They retain all the evidence that e-BH uses, and the difference $e_j-h_j(1,X)\geq0$ is the evidence that it discards.

	Conditional calibration spends this discarded evidence. Suppose that under $H_j$ a statistic $S_j$ makes the conditional law of $X$ given $S_j$ known and common to all distributions in $H_j$. Then
	\[
	\Phi_j(c;S_j)=\E[h_j(c,X)-e_j(X)\mid S_j]
	\]
	can be computed, for instance by sampling from this conditional law. It is non-decreasing in $c$ and nonpositive at $c=1$, so the inflation can be raised until $\Phi_j$ reaches zero. For an $S_j$-measurable $c_j\geq1$ with $\Phi_j(c_j;S_j)\leq0$, the recalibrated value $e'_j=h_j(c_j,X)$ satisfies
	\[
	\E[e'_j\mid S_j]=\E[e_j\mid S_j]+\Phi_j(c_j;S_j)\leq\E[e_j\mid S_j],
	\]
	because $c_j$ is fixed once $S_j$ is known, and it is therefore an e-value under $H_j$. For $j\in R_0$, $c_je_j\geq e_j\geq t_j$, so $e'_j=K/(\alpha r)$, and e-BH applied to $e'$ rejects at least $R_0$. A hypothesis outside $R_0$ is added whenever its inflated e-value reaches $K/(\alpha(r+1))$.

	This is exactly Proposition~\ref{prop:conditional} in the language of Section~\ref{subsec:recert}. The conditioning statistic is $S_j$, the baseline is the pruned e-value $h_j(1,X)$ with conditional mean $H=\E[h_j(1,X)\mid S_j]$, and the budget is $B=\E[e_j\mid S_j]$, whose mean is at most one under $H_j$, although it may exceed one for some values of $S_j$. The increment $W=h_j(c_j,X)-h_j(1,X)$ is nonnegative, and since $B-H=\E[e_j-h_j(1,X)\mid S_j]$, the condition $\Phi_j(c_j;S_j)\leq0$ is exactly \eqref{eq:conditional-increment}. The recalibrated e-values then certify rejection sets through the e-collection \eqref{eq:ebh-collection} evaluated at $e'$, whose largest certified set is $\eBH_\alpha(e')$. Conditional calibration thus spends unused budget across the values of the data that share the same $S_j$, whereas RU-eBH spends it across the values of the external seed.

	Finally, \citet{leeren2024} take the largest admissible factor $c^*_j=\sup\{c\geq1:\Phi_j(c;S_j)\leq0\}$, which need not be attained, because $h_j(c,X)$ jumps in $c$. If $\Phi_j(c^*_j;S_j)>0$, the value $t_j(X)\ind\{c^*_je_j(X)>t_j(X)\}$, with strict inequality, is the increasing limit of $h_j(c,X)$ as $c\uparrow c^*_j$, so monotone convergence preserves the bound on its conditional mean, and since $c^*_j>1$, it still equals $K/(\alpha r)$ on $R_0$.

\end{document}